\documentclass[usenatbib]{mn2e}
\usepackage{wrapfig}
\usepackage{sidecap}
\usepackage{graphicx}
\usepackage{caption}
\usepackage{longtable}
\usepackage{array}
\usepackage{subcaption}
\usepackage[bookmarks, colorlinks, breaklinks, pdftitle={FQXi Essay},pdfauthor={Author}]{hyperref}  
\hypersetup{linkcolor=blue,citecolor=blue,filecolor=black,urlcolor=blue}

\title[Gas content \& interaction as the drivers of kinematic asymmetry in SAMI galaxies]{{The SAMI Galaxy Survey: gas content and interaction as the drivers of kinematic asymmetry}}

\author[Bloom et al.]
{\parbox{\textwidth} 
{J.~V.~Bloom$^{1,2}$,
S.~M.~Croom$^{1,2}$,
J.~J.~Bryant$^{1,2,3}$,
A.~L.~Schaefer$^{1,2,3}$,
J.~Bland-Hawthorn$^{1}$,
S.~Brough$^{2,4}$,
J.~Callingham$^{1,2,5}$,
L.~Cortese$^{6}$,
C.~Federrath$^{7}$,
N.~Scott$^{1,2}$,
J.~van~de~Sande$^{1}$,
F.~D'Eugenio$^{7}$,
S.~Sweet$^{8}$,
C.~Tonini$^{8}$,
J.~T.~Allen$^{1}$,
M.~Goodwin$^{3}$,
A.~W.~Green$^{3}$,
I.~S.~Konstantopoulos$^{2,9}$,
J.~Lawrence$^{3}$,
N.~Lorente$^{3}$,
A.~M.~Medling$^{7,10}$,
M.~S.~Owers$^{3,11}$,
S.~N.~Richards$^{1,2,3}$,
R.~Sharp$^{7}$
}
\vspace{0.4cm} \\
\parbox{\textwidth}{$^{1}$Sydney Institute for Astronomy, School of Physics, University of Sydney, NSW 2006, Australia\\
$^{2}$CAASTRO: ARC Centre of Excellence for All-sky Astrophysics\\
$^{3}$Australian Astronomical Observatory (AAO), 105 Delhi Rd, North Ryde, NSW 2113, Australia\\
$^{4}$School of Physics, University of New South Wales, NSW 2052, Australia\\
$^{5}$CSIRO Astronomy \& Space Science P.O.Box 76, Epping, NSW 1710, Australia\\
$^{6}$International Centre for Radio Astronomy Research, University of Western Australia, 35 Stirling Highway, Crawley, WA, 6009, Australia\\
$^{7}$Research School of Astronomy and Astrophysics, Australian National University, Canberra, ACT 2611, Australia\\
$^{8}$Centre for Astrophysics and Supercomputing, Swinburne University of Technology, PO Box 218, Hawthorn, VIC 3122, Australia\\
$^{9}$Atlassian 341 George St Sydney, NSW 2000\\
$^{10}$Cahill Center for Astronomy and Astrophysics, California Institute of Technology, MS 249-17 Pasadena, CA 91125, USA\\
$^{11}$Department of Physics and Astronomy, Macquarie University, NSW 2109, Australia}
}

\begin{document}

\maketitle

\begin{abstract}
In order to determine the causes of kinematic asymmetry in the H$\alpha$ gas in the SAMI Galaxy Survey sample, we investigate the  comparative influences of environment and intrinsic properties of galaxies on perturbation. We use spatially resolved H$\alpha$ velocity fields from the SAMI Galaxy Survey to quantify kinematic asymmetry ($\overline{v_{asym}}$) in nearby galaxies and environmental and stellar mass data from the GAMA survey. 

{We find that local environment, measured as distance to nearest neighbour, is inversely correlated with kinematic asymmetry for galaxies with  $\mathrm{\log(M_*/M_\odot)}>10.0$, but there is no significant correlation for galaxies with  $\mathrm{\log(M_*/M_\odot)}<10.0$.    Moreover, low mass galaxies ($\mathrm{\log(M_*/M_\odot)}<9.0$) have greater kinematic asymmetry at all separations, suggesting a different physical source of asymmetry is important in low mass galaxies.}

We propose that secular effects derived from gas fraction and gas mass may be the primary causes of asymmetry in low mass galaxies. High gas fraction is linked to high $\frac{\sigma_{m}}{V}$ (where $\sigma_m$ is H$\alpha$ velocity dispersion and $V$ the rotation velocity), which is strongly correlated with $\overline{v_{asym}}$, and galaxies with $\log(M_*/M_\odot)<9.0$ have offset $\overline{\frac{\sigma_{m}}{V}}$ from the rest of the sample. Further, asymmetry as a fraction of dispersion decreases for galaxies with $\log(M_*/M_\odot)<9.0$. Gas mass and asymmetry are also inversely correlated in our sample. We propose that low gas masses in dwarf galaxies may lead to asymmetric distribution of gas clouds, leading to increased relative turbulence.





\end{abstract}


\section{Introduction}

Dwarf galaxies are known to be largely irregular systems  [e.g. \citet{van1998neutral,cannon2004complex,lelli2014dynamics}] with low luminosity. They also typically have higher gas fractions than their larger counterparts \citep{geha2006baryon,huang2012gas}, and contain comparatively large disturbances, both as a result of interactions and of stochastic processes \citep{escala2008stability}. Complex H$\alpha$ kinematics in low stellar mass galaxies are linked to the presence of multiple star-forming clumps \citep{amorin2012complex}. Further, {dwarfs} have been found to have higher turbulent support, compared to rotational support \citep{barton2001tully,garrido2005ghasp}, as well as irregular morphologies  \citep{roberts1994physical,mahajan2015galaxy}. 

One of the primary sources of kinematic disturbance is interaction between galaxies. The formation of dark matter halos through a series of mergers is also well-established in $\Lambda$CDM cosmology, both from theory and observation \citep{peebles1982large,cole2008statistical,neistein2008merger}. N-body simulations suggest that the rate of galaxy and halo mergers should be dependent on redshift, but there are differences in merger rate between the predictions of theory and the results of observation. Indeed, simulated major merger rates are highly variable \citep{hopkins2010mergers}. As a result, innovative methods of studying the merger rate and influence on galaxy evolution are an important part of understanding the formation of the modern universe. In particular, observations of kinematics are needed to probe beyond morphology.

Integral field spectroscopy (IFS) surveys such as SINS  \citep{shapiro2008kinemetry} and ATLAS\textsuperscript{3D} \citep{cappellari2011atlas3d}  have used the 2D kinematics of galaxies to find and quantify disturbances in galaxies at high and low redshift. They are, however, limited in scope by their use of monolithic instruments, which requires taking IFS measurements for one object at a time, thus increasing the time necessary to accrue large sample sizes. The Sydney--AAO Multi-object IFS (SAMI) is a multiplexed spectrograph, that allows for sample sizes of thousands of galaxies on a much shorter timescale than is possible with other instruments \citep{croom2012sydney}. The SAMI Galaxy Survey sample will contain 3600 galaxies selected from a combination of the Galaxy And Mass Assembly (GAMA) survey sample \citep{driver2011galaxy} and from 8 galaxy clusters \citep{bryant2015sami,owers2017sami}. The chosen GAMA galaxies reflect a broad range in stellar mass and environment, in accordance with the science drivers of the SAMI Galaxy Survey \citep{bryant2015sami}.

Previous large-scale surveys have used images to study disturbance and merger rates [e.g. \citet{casteels2013galaxy,conselice2003evidence}]. This work has included techniques such as the counting of close pairs and the Concentration, Asymmetry, Smoothness (CAS) classification scheme \citep{conselice2003evidence}. Image-based analysis has been used to determine merger rates in the nearby and high redshift universe \citep{lotz2011major}. However, there are inherent difficulties arising from the influence of Hubble type, and these studies are restricted in the range of perturbations to which they are sensitive  \citep{conselice2003evidence}. For example, at high redshift, galaxies may appear photometrically perturbed, but remain kinematically regular, due to features such as clumped star formation \citep{shapiro2008kinemetry}. Kinematics provide a different perspective on galaxies' internal processes, and can be used to probe these processes without bias due to Hubble type or morphological features.

In previous work \citep{bloom2016sami}, we developed a quantitative measure of asymmetry in gas dynamics, $\overline{v_{asym}}$, based on kinemetry and following the method of \citet{shapiro2008kinemetry}. We found a strong inverse correlation between stellar mass and $\overline{v_{asym}}$. Here, we investigate this relationship further, determining how environment is related to asymmetry. 


In Section \ref{sec:sami}, we briefly review the SAMI instrument and the SAMI Galaxy Survey data used in this work. In Section~\ref{sec:data} we describe the GAMA Survey data used here and introduce the kinemetry algorithm as a means of measuring kinematic asymmetry. In Section \ref{sec:results}, we show the relationships between environment, asymmetry and stellar mass in our sample. Section \ref{sec:discussion} contains a discussion of environmental factors and other causes of kinematic asymmetry. We conclude in Section \ref{sec:conclusion}. We assume a standard cosmology, with $\Omega_m = 0.3,\ \Omega_\lambda = 0.7$ and H$_0 = 70$ km/s/Mpc.

\section{The SAMI Galaxy Survey instrument and sample}
\label{sec:sami}

The SAMI instrument is a multiplexed integral field unit on the Anglo-Australian Telescope, which uses imaging fibres, or hexabundles  \citep{bland2011hexabundles,bryant2014sami}, to simultaneously produce integral field spectra for multiple objects \citep{bryant2011characterization}. Each SAMI hexabundle consists of 61 optical fibres, with a total bundle diameter of 15 arcseconds. The instrument is installed with a fibre cable feeding to the double-beamed AAOmega spectrograph  \citep{sharp2006performance,croom2012sydney}.

The 3600 galaxies comprising the SAMI Galaxy Survey sample were chosen from the GAMA survey sample \citep{driver2009gama} and 8 galaxy clusters forming an additional part of the sample. The galaxies were selected from within the GAMA sample to represent a broad range in environment density and stellar mass. The complete SAMI Galaxy Survey sample contains four volume-limited sub-samples and supplementary, low redshift dwarf galaxies \citep{bryant2015sami}. 

The key data products from the SAMI Galaxy Survey data reduction pipeline are datacubes, from which emission line maps are derived. A full description of the SAMI Galaxy Survey data reduction pipeline can be found in  \citet{sharp2015sami}.

The current paper builds on the work of \citet{bloom2016sami}, and uses similar data products from the SAMI Galaxy Survey, including {\small LZIFU}{\sc}. {\small LZIFU}{\sc} is a spectral fitting pipeline written in {\small IDL}{\sc} (Interactive Data Language), performing flexible emission line fitting in the SAMI Galaxy Survey datacubes. The {\small LZIFU}{\sc} pipeline products are 2D kinematic and emission line strength maps for user-assigned lines. A more complete description of the {\small LZIFU}{\sc} pipeline can be found in \citet{ho2016sami}.

In this work, we used SAMI Galaxy Survey internal data release v.0.9.1. This provided 1021 galaxies that had been observed and processed through the  {\small LZIFU}{\sc} pipeline. In previous work, we used an H$\alpha$ signal to noise (S/N) cut of 10 to exclude noisy spaxels from the velocity fields output by  {\small LZIFU}{\sc}. Here, we relax the S/N cut to 6. The median velocity error did not increase, so we found that this increased the number of galaxies in the sample without compromising data quality. 827 galaxies met the S/N cut requirements and yielded results from kinemetry, a 230$\%$ increase on the sample size in  \citet{bloom2016sami}. Note that we do not use cluster galaxies in this sample.

\section{Input data}
\label{sec:data}
\subsection{Environment and stellar mass}

This work uses data from the GAMA Survey. Environment measurements are taken from the G3CGal Groups catalogue \citep{robotham2011galaxy}, the Environment Measures catalogue \citep{brough2013galaxy}, and distances to the first nearest neighbour. From the Environment Measures catalogue we take the distance to fifth nearest neighbour, $d_5$. The $d_5$ calculations are done on a density defining pseudo-volume-limited population of galaxies in the GAMA sample. The first nearest neighbour distances, $d_1$ are calculated similarly to \citet{robotham2014galaxy}. They are projected distances between GAMA galaxies. The maximum allowed separation is 5 Mpc or the distance to the nearest sample edge, whichever is closest, and the allowable redshift offset between paired galaxies is $\pm1000$km/s. The mean redshift of the two galaxies is the redshift at which the separation is calculated.

The stellar mass measurements are from the GAMA Survey catalogue StellarMasses \citep{taylor2011galaxy}, and are based on $ugriz$ SEDs constructed from matched aperture photometry. Typical errors are {0.1 dex}.

\subsection{Kinematic asymmetry}
\label{sec:kinemetry}
Kinemetry is a generalisation of photometry to the higher order moments of the line of sight velocity distribution \citep{krajnovic2006kinemetry}. {The kinemetry code is written in IDL, and can be found at \url{http://davor.krajnovic.org/idl/}. It was used, {for example,} to map and quantify asymmetries in the stellar velocity (and velocity dispersion) maps of galaxies in the ATLAS\textsuperscript{3D} survey {\citep{krajnovic2011atlas3d}}. 

{The algorithm models} kinematic maps as a sequence of concentric ellipses, with parameters defined by the { galaxy kinematic position angle (PA), centre and ellipticity.} It is possible to fit the latter two parameters within kinemetry, or to determine them by other means and exclude them from the fitting procedure. For each ellipse, the kinematic moment is extracted and decomposed into the Fourier series:
\begin{equation}
K(a,\psi)=A_{0}(a)+\sum_{n=1}^{N}(A_{n}(a)sin(n\psi)+B_{n}(a)cos(n\psi)),
\label{equation:orig}
\end{equation}
where $\psi$ is the azimuthal angle in the galaxy plane, and $a$ is the ellipse semi-major axis length. Points along the ellipse perimeter are sampled uniformly in $\psi$, making them equidistant in circular projection. The series can be expressed more compactly, as  \citep{krajnovic2006kinemetry}:
\begin{equation}
K(a,\psi)=A_{0}(a)+\sum_{n=1}^{N}k_{n}(a)cos[n(\psi-\phi_{n}(a))],
\end{equation}
with the amplitude and phase coefficients ($k_{n},\phi_{n}$) defined as:
\begin{equation}
k_{n}=\sqrt{A^{2}_{n}+B^{2}_{n}}
\end{equation}
and
\begin{equation}\label{eq:phi}
\phi_{n}=arctan\left(\frac{A_{n}}{B_{n}}\right) .
\end{equation}

The moment maps for both velocity and velocity dispersion are thus described by the geometry of the ellipses and power in the coefficients $k_{n}$ of the Fourier terms as a function of \textit{a} \citep{krajnovic2006kinemetry}. The velocity field of a perfect disc, with no asymmetries, would be entirely expressed by the $cos(\psi)$ term {of Equation~\ref{equation:orig}}, with zero power in the higher order modes, since the velocity peaks at the galaxy major axis and goes to zero along the minor axis. Accordingly, the power in the $B_{1}$ term represents circular motion, with deviations carried in the other coefficients. Analogously, a map of the velocity dispersion field of a perfectly regular, rotating disc would have all power in the $A_{0}$ term (i.e. radial dispersion gradient) \citep{krajnovic2006kinemetry}. The velocity dispersion field is an even moment of the LOSVD, and thus its kinemetric analysis reduces to traditional surface photometry. 

In \citet{bloom2016sami} we defined a quantitative metric, $\overline{v_{asym}}$, to parametrise kinematic asymmetry in the SAMI Galaxy Survey sample, derived from kinemetry \citep{krajnovic2006kinemetry,shapiro2008kinemetry} and this was validated by comparison to a thorough by-eye classification based on SDSS imaging.
\begin{figure}
\centering
\includegraphics[width=9cm]{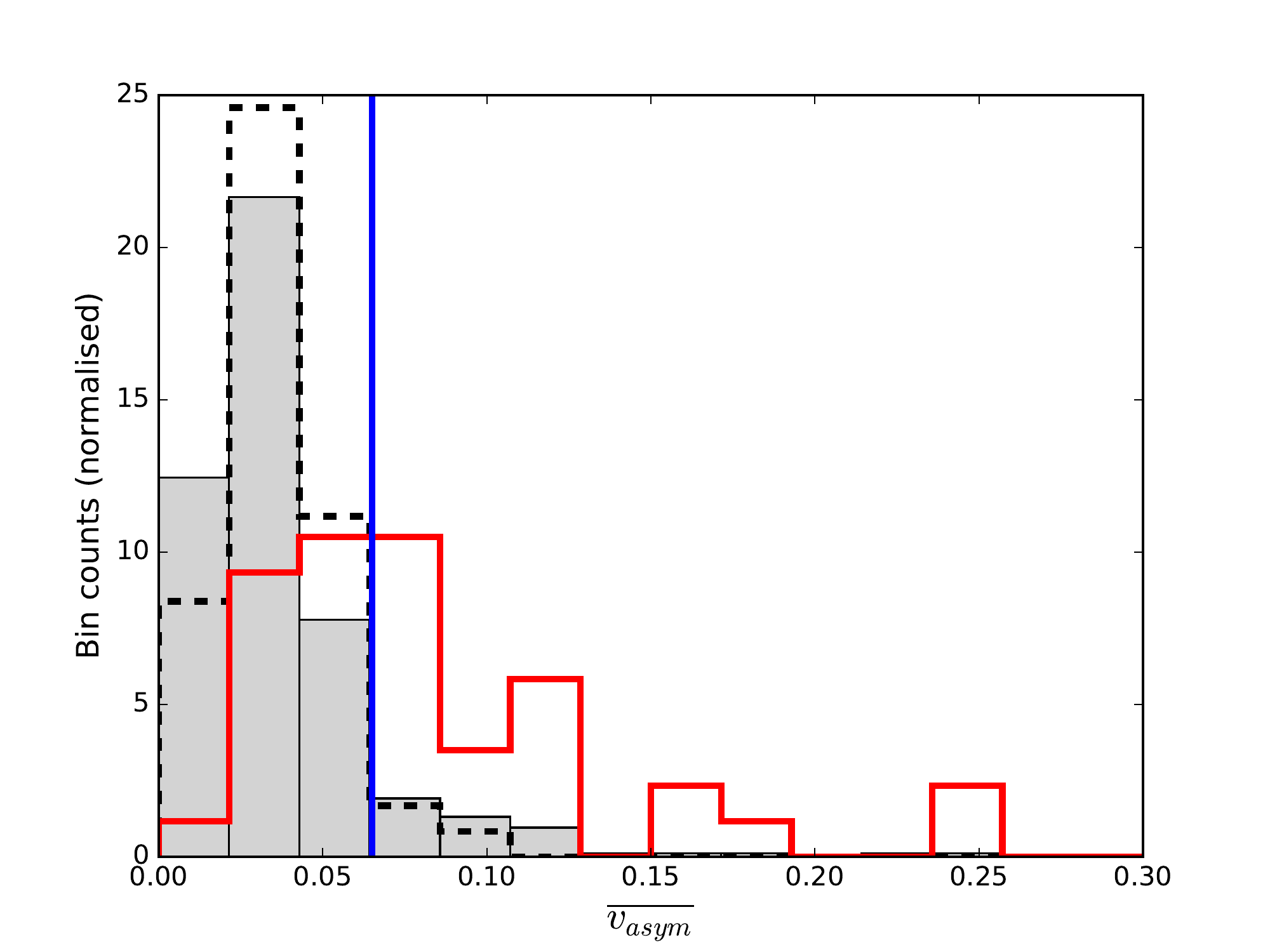}
\caption{Histogram of $\overline{v_{asym}}$ for the full sample in this work (grey) and galaxies classified as normal (black, dashed) and perturbed (red) in previous work. The asymmetry cutoff from  \citet{bloom2016sami} is shown in blue. Visually classified asymmetric and normal galaxies are seen to have offset distributions of $\overline{v_{asym}}$, as in \citet{bloom2016sami}.}
\label{fig:asym_hist}
\end{figure}

The kinemetry algorithm fits a series of regularly spaced, concentric ellipses, from which radial moment values are taken. The ellipses are fit with parameters defined by the galaxy centre, kinematic position angle (PA), and ellipticity. Ellipticities are calculated using 400\arcsec $r$-band cutout images from SDSS. The input image is processed with SExtractor to mask out neighbouring objects. PSFEx is used to build a model PSF at the location of the galaxy centre. The ellipticity is then found as a light-weighted mean for the whole galaxy, using the Multi-Gaussian Expansion (MGE) technique \citep{emsellem1994multi} and code from \citet{cappellari2002efficient}. For more detail, we refer to D'Eugenio et al (in prep.). We slightly alter the method in \citet{bloom2016sami} and use the kinematic, rather than photometric PA. We do this because we wish to isolate the effects of misalignment between the photometric and kinematic PA from other forms of atypical kinematics (see Section~\ref{sec:int}). The kinematic PA is calculated by the FIT\_KINEMATIC\_PA routine from \citet{krajnovic2006kinemetry}. The combination of the fitted ellipses is used to construct individual moment maps. As in \citet{bloom2016sami}, and following \citet{shapiro2008kinemetry}, 
\begin{equation}
\overline{v_{asym}}=\overline{\left({\frac{k_3+k_5}{2k_1}}\right)}
\end{equation}
where $k_1, k_3, k_5$ are the coefficients of the Fourier decomposition of the first, third and fifth order moments of the line of sight velocity distribution, respectively. Regular rotation is carried in the $k_1$ term, and asymmetry in the odd higher order terms. We do not use the even higher order modes because these carry asymmetry in the velocity dispersion. 

 Fig.~\ref{fig:asym_hist} shows the distribution of  $\overline{v_{asym}}$ for the full sample in this work (grey), with galaxies visually classified in \citet{bloom2016sami} as normal (black, dashed) and perturbed (red) and the asymmetry cutoff in \citet{bloom2016sami}. Note that the sample in \citet{bloom2016sami} was smaller than that used in this work, so the galaxies classified as normal and asymmetric in Fig.~\ref{fig:asym_hist} do not comprise the full sample used here. Visually asymmetric and normal galaxies are seen to have offset distributions of $\overline{v_{asym}}$.  {We note that we also make use of this metric in \citet{bloom2017sami}, in an identical manner to that described above.}

\section{Asymmetry vs. stellar mass and distance to nearest neighbour}
\label{sec:results}


\subsection{Distance to fifth nearest neighbour and $\overline{v_{asym}}$}

Fig.~\ref{fig:5nn_asym} shows the comoving distance to a galaxy's fifth nearest neighbour (in Mpc) \citep{brough2013galaxy}, henceforth $d_5$, against $\overline{v_{asym}}$. We find no relationship between $\overline{v_{asym}}$ and $d_5$. A Spearman rank correlation test of $\overline{v_{asym}}$ and $d_5$ gives $\rho=0.039, p=0.28$. We define a cutoff distance of $1 Mpc$ to delineate galaxies `far' from and `near' to their fifth nearest neighbour. The median $\overline{v_{asym}}$ for far galaxies is $0.040\pm0.0001$, and for near galaxies it is $0.039\pm0.0005$. That is, there is no offset in the median of the two distributions. These results do not change for different distance cutoffs, or for different ranges of stellar mass.

\begin{figure}
\centering
\includegraphics[width=9cm]{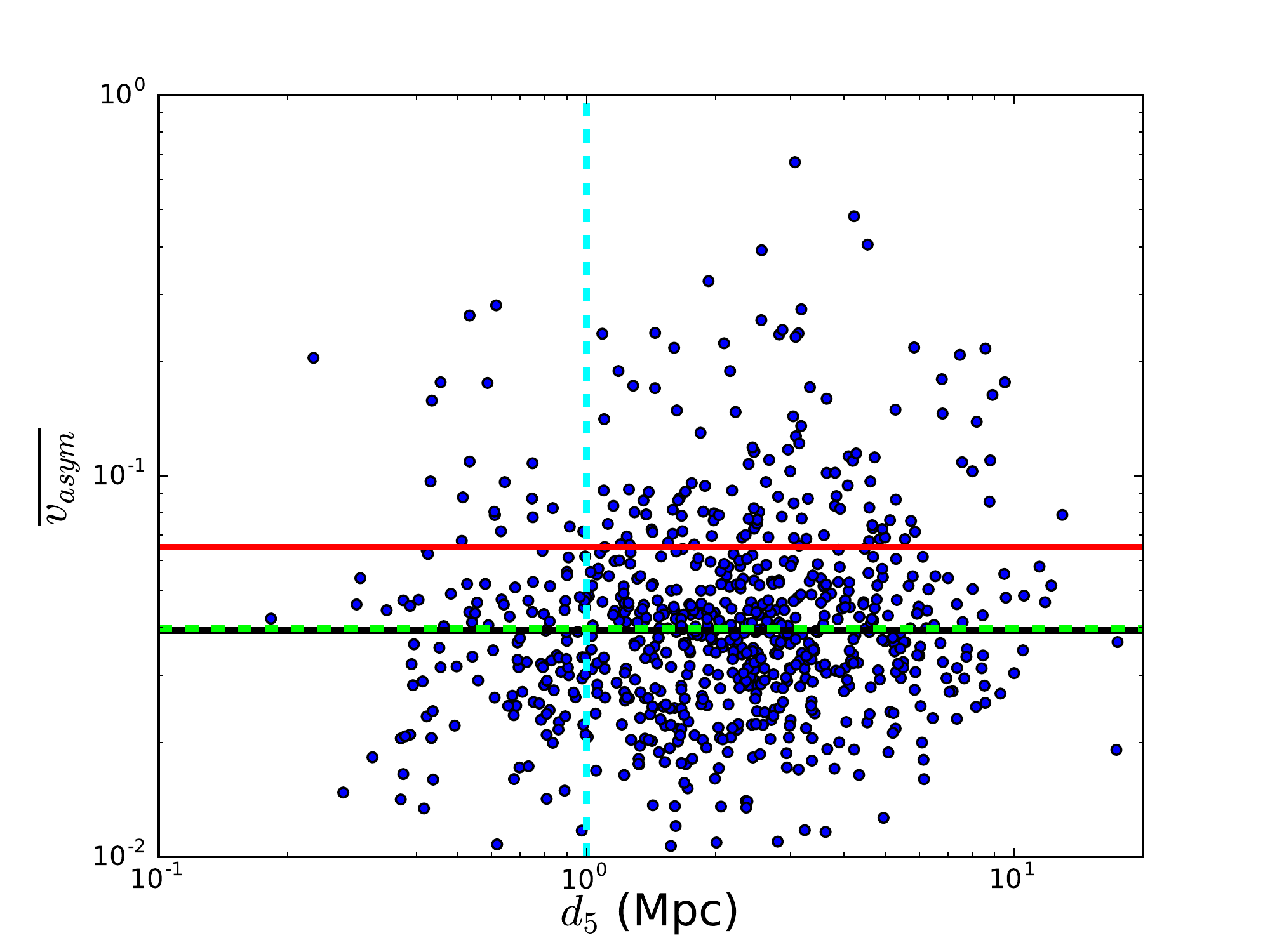}
\caption{$\overline{v_{asym}}$ against $d_5$, the comoving distance between a galaxy and its fifth nearest neighbour. The cyan line is at 1Mpc, distinguishing between galaxies close to and far from their fifth nearest neighbour. The red line shows the asymmetry cutoff at $\overline{v_{asym}}=0.065$, from \citet{bloom2016sami}. The black and green lines are the median $\overline{v_{asym}}$ for galaxies in dense and sparse environments, respectively, and there is no significant offset between them. We conclude that there is no correlation between the fifth nearest neighbour distance and the kinematic asymmetry.}
\label{fig:5nn_asym}
\end{figure}

\subsection{First nearest neighbour, stellar mass and $\overline{v_{asym}}$}
\label{sec:dn}

Close pairs are frequently used as a proxy indicator of mergers [e.g. \citet{darg2010galaxy}, \citet{ellison2013galaxy}, \citet{robotham2014galaxy}, \citet{barrera2015central}]. Accordingly, we measure the distance to the first nearest neighbour, with close distances being assumed to entail interaction. 

Fig.~\ref{fig:dist_asym_sm} shows distance to nearest neighbour against asymmetry. We colour the points by stellar mass, and see that there is an inverse correlation between stellar mass and asymmetry, confirming the results of \citet{bloom2016sami}, with a Spearman rank correlation test giving $\rho=-0.23, p=1\times10^{-7}$. There is a weak inverse correlation between $\overline{v_{asym}}$ and $d_{1}$ for galaxies with $\log(M_{*}/M_\odot)>10.0$, with $\rho=-0.19, p=7\times10^{-4}$. For galaxies with $\log(M_{*}/M_\odot)<10.0$, there is no significant correlation, with $\rho=0.016, p=0.74$. For a more detailed exploration of the `transition' mass at which a correlation is significant, see Appendix \ref{sec:appendix}.  

Fig.~\ref{fig:dist_asym_sm} also shows median asymmetry for low and high stellar mass galaxies, in bins of separation. This isolates the effects of environment and stellar mass by splitting the populations. Across the sample, low stellar mass galaxies have higher median asymmetry, regardless of distance to nearest neighbour. The median difference across all stellar mass bins is  0.018 $\pm$ 0.003. {Fig.~\ref{fig:asym_hists_d1} shows histograms of $\overline{v_{asym}}$ for different stellar mass populations across two ranges of $d_1$, chosen to be less than and greater than the median of $d_1$, respectively. In both bins of $d_1$, galaxies with $\log(M_{*}/M_\odot)<9.0$ have a distribution of $\overline{v_{asym}}$ shifted towards higher values. We note that the following results do not change when different values are chosen as a cutoff for $d_1$.}

A two-sample Kolmogorov-Smirnov test of the distributions of asymmetries of galaxies with $\log(M_{*}/M_\odot)<10.0$ and those with $\log(M_{*}/M_\odot)>10.0$ gives $p=4\times10^{-5}$. {When the sample is divided into bins of $d_1$, as in Fig.~\ref{fig:asym_hists_d1}, similar results are produced from separate two-sample Kolmogorov-Smirnov tests. When testing the distributions of $\overline{v_{asym}}$ for high mass galaxies across the $d_1$ bins in Fig.~\ref{fig:asym_hists_d1}, the same test shows that the distributions are different ($p=4\times10^{-2}$, i.e. marginal significance). However, this is not the case for low mass galaxies ($p=6\times10^{-1}$).} We conclude that there are global differences in the distribution of asymmetries for high and low mass galaxies. Fig.~\ref{fig:dist_bins} shows these patterns more clearly, with $d_1$ against $\overline{v_{asym}}$ in bins of stellar mass. As the scatter in the relationship in the $\log(M_*/M_\odot)>10.0$ panel is large, we also show the median $d_1$ for galaxies with $\overline{v_{asym}}>0.065$ and $\overline{v_{asym}}>0.065$ for all three subsamples. There is a large offset ($71\pm39$kpc) between these medians in the $\log(M_*/M_\odot)>10.0$ sample, but not in the lower mass cases.

{There is considerable scatter in the distributions of $\overline{v_{asym}}$ against $d_1$. Fig.~\ref{fig:rms} shows the root mean square (RMS) of the distribution of $\log(\overline{v_{asym}})$ in bins of $d_1$, for high and low mass galaxies.  Errors are produced by bootstrapping 100 iterations of adding random noise (based on uncertainties in $\overline{v_{asym}}$) to the distribution of $\overline{v_{asym}}$ within each mass bin. At all $d_1$, the scatter in the distribution of $\overline{v_{asym}}$ of galaxies with $\log(M_{*}/M_\odot)<9.0$ is larger than for galaxies with $\log(M_{*}/M_\odot)>10.0$. At low $d_1$, there is an increase in scatter for high mass galaxies, but only weak evidence of this effect in the low mass population.  } {Despite this difference in significance, the overall trends in the two populations look similar. A larger sample may thus result in a significant result for the low mass galaxies as well.}

{Fig.~\ref{fig:fracs} shows $\overline{v_{asym}}$ against $d_1$, with the fractional count of galaxies with $\overline{v_{asym}}>0.065$ for high and low mass populations, respectively. There is no significant decrease in fractional count with increased $d_1$ for either population of galaxies.}


In order to test whether the calculated correlation in the $\log(M_{*}/M_\odot)>10.0$ case {of Fig.~\ref{fig:dist_bins}} is significant, we resampled the $\overline{v_{asym}}$ values within Gaussian distributions centred on the calculated value, with width equal to the error and then recalculated the correlation coefficients, over 1000 iterations. We found that, for all 1000 cases, $-0.25<\rho<-0.13$ and $p<0.018$. Further, 60\% of $p$-values were $<3\sigma$. The same test on the $\log(M_*/M_\odot)<9.0$ data gave ranges of $-0.1<\rho<0.5$, $0.3<p<1.0$. Thus, whilst the trend is of equal size to the scatter, it is still significant. 





\begin{figure}
\centering
\includegraphics[width=9cm]{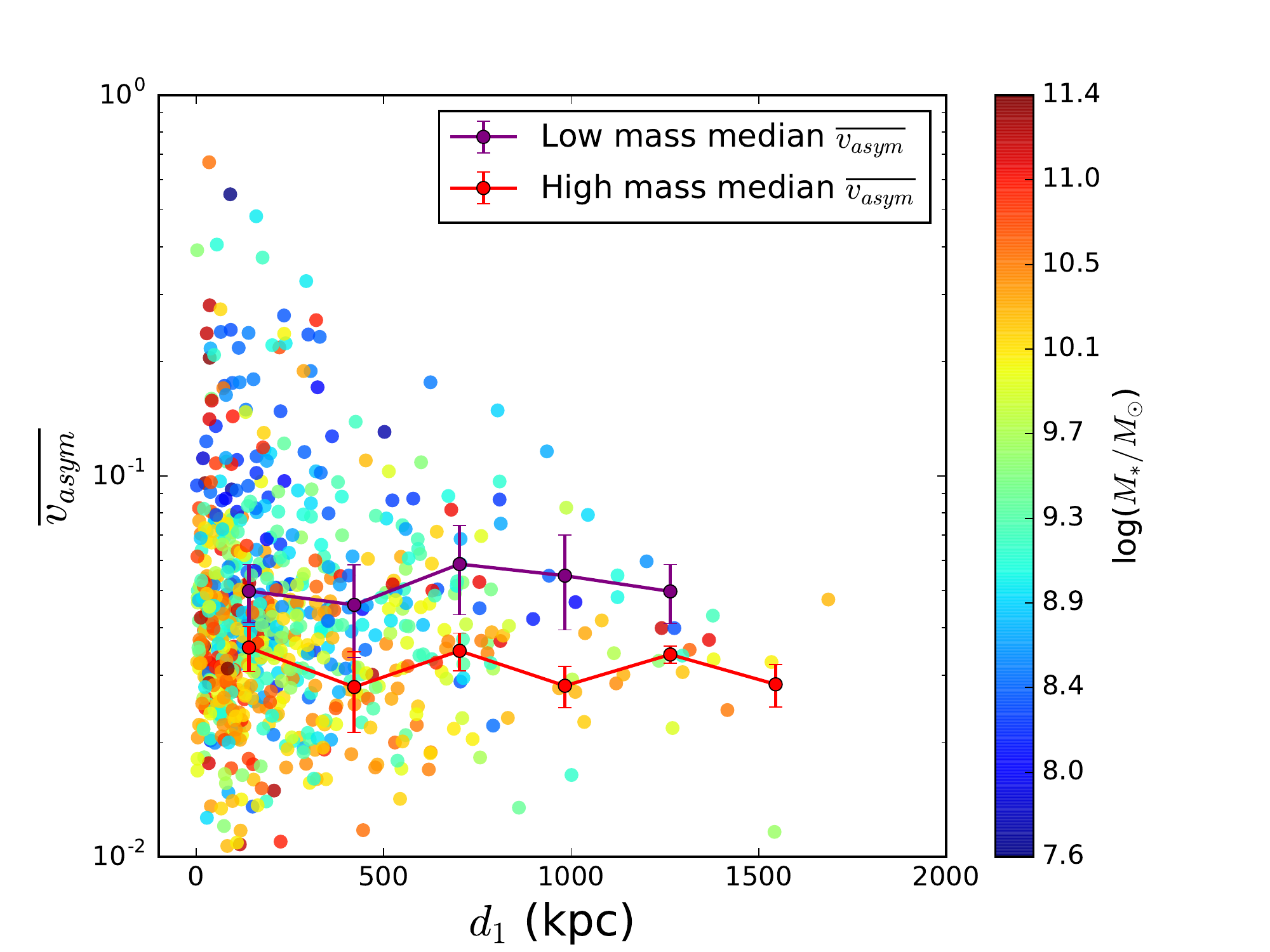}
\caption{$\overline{v_{asym}}$ against the comoving distance between a galaxy and its first nearest neighbour, $d_1$, with points coloured by stellar mass. There is an inverse correlation between stellar mass and $d_1$ , as well as stellar mass and $\overline{v_{asym}}$. There is a weak inverse correlation between $d_1$ and $\overline{v_{asym}}$ for galaxies with $\log(M_*/M_\odot) > 10.0$, but not for the rest of the sample. The connected points show median asymmetry for galaxies with $\log(M_*/M_\odot) > 10.0$ (red) and $\log(M_*/M_\odot) < 9.0$ (purple), in bins of separation. In all cases, low stellar mass galaxies have higher median asymmetry.}
\label{fig:dist_asym_sm}
\end{figure}

\begin{figure*}
\centering
\includegraphics[width=\textwidth]{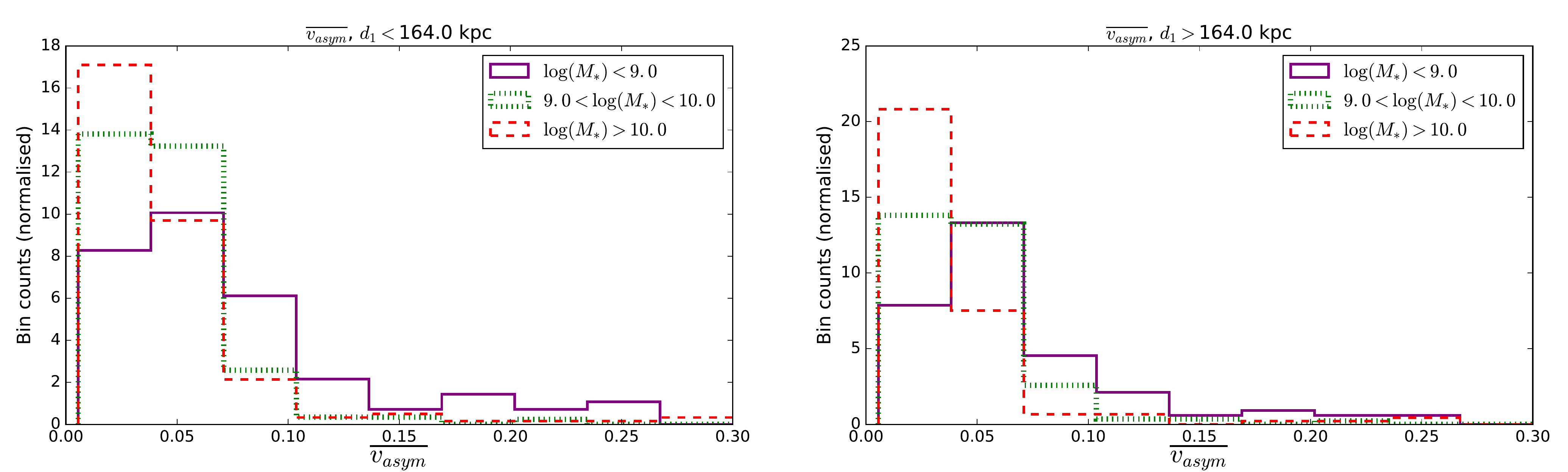}
\caption{{Histograms of $\overline{v_{asym}}$, in bins of stellar mass (dashed, dotted and solid lines for $\log(M_{*}/M_\odot)>10.0$, $9.0<\log(M_{*}/M_\odot)<10.0$ and $\log(M_{*}/M_\odot)<9.0$, respectively) across two ranges of $d_1$. In both $d_1$ ranges, galaxies with $\log(M_{*}/M_\odot)<9.0$ have higher distributions of $\overline{v_{asym}}$. }}
\label{fig:asym_hists_d1}
\end{figure*}

\begin{figure*}
\centering
\includegraphics[width=\textwidth]{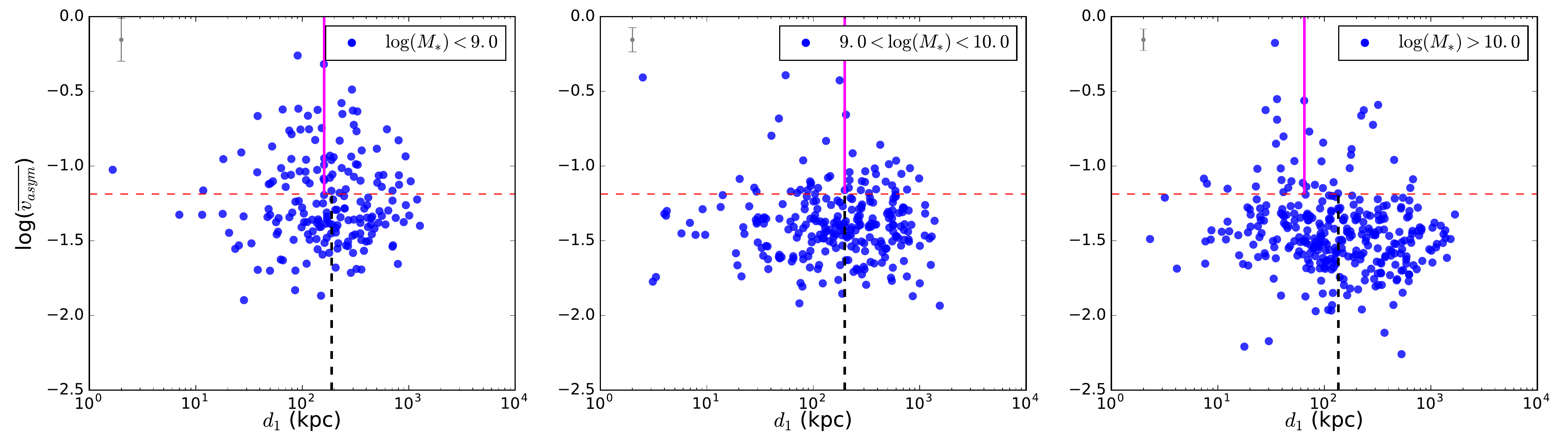}
\caption{$d_1$ against $\overline{v_{asym}}$, in bins of stellar mass. There is a weak inverse correlation between $d_{1}$ and $\overline{v_{asym}}$ for galaxies with $\log(M_{*}/M_\odot)>10.0$, but not for the two lower stellar mass bins. Low and high stellar mass galaxies also have different distributions of $\overline{v_{asym}}$ and $d_1$. Typical error bars are shown in grey (top left corner), and the $\overline{v_{asym}}>0.065$ cutoff from \citet{bloom2016sami} is shown in red (dashed). The median $d_1$ for galaxies with $\overline{v_{asym}}>0.065$ (pink, $65\pm33$kpc) and $\overline{v_{asym}}>0.065$ (black, dashed, $135\pm22$kpc) is shown for all three subsamples. There is a statistically significant offset of $71\pm39$kpc between these medians in the $\log(M_*/M_\odot)>10.0$ sample, but not in the lower mass cases.}
\label{fig:dist_bins}
\end{figure*}

\begin{figure}
\centering
\includegraphics[width=9cm]{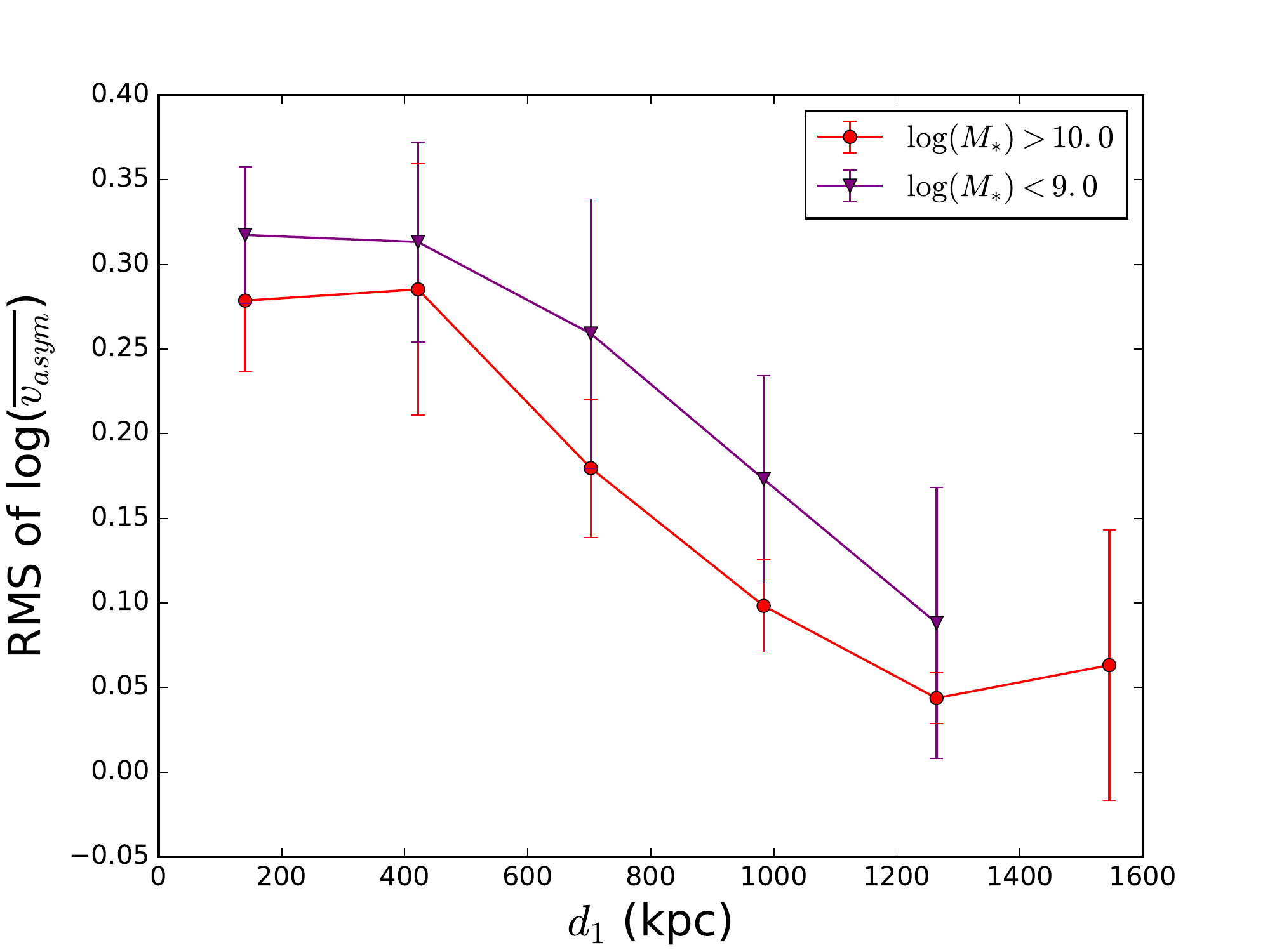}
\caption{{RMS of the distribution of $\log(\overline{v_{asym}})$ in bins of $d_1$, for both high and low mass galaxies. At all $d_1$, there is more scatter in the distribution of galaxies with $\log(M_{*}/M_\odot)<9.0$  than for those with $\log(M_{*}/M_\odot)>10.0$. At low $d_1$, there is an increase in the RMS of the distribution of $\overline{v_{asym}}$ for high mass galaxies, but low mass galaxies show only a weak change in distribution. }}
\label{fig:rms}
\end{figure}

\begin{figure}
\centering
\includegraphics[width=8.2cm]{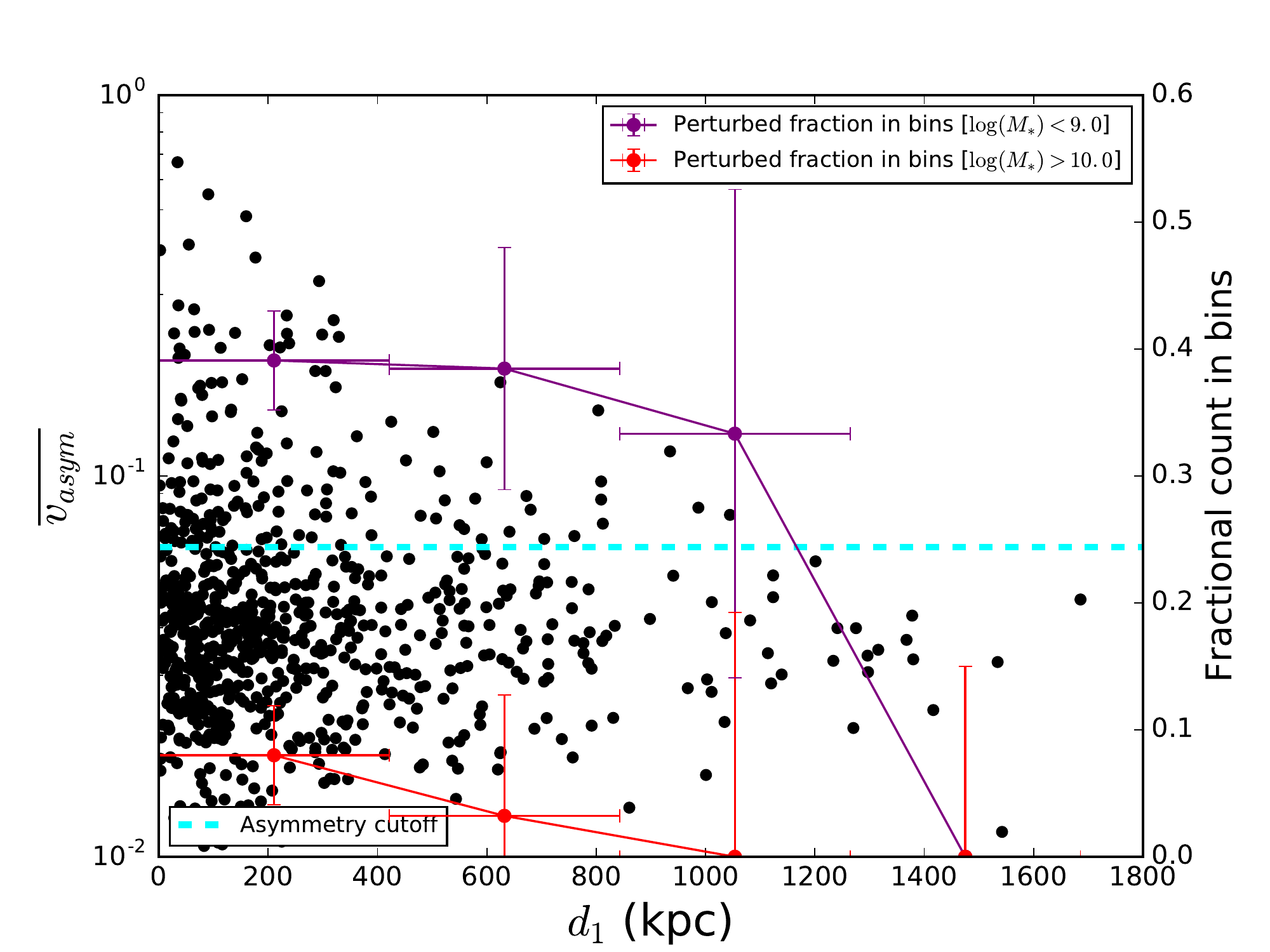}
\caption{{$\overline{v_{asym}}$ against $d_1$, with the fractional count of galaxies with $\overline{v_{asym}}>0.065$ for high (red) and low (purple) mass populations. The asymmetry cutoff at $\overline{v_{asym}}=0.065$ is shown in cyan (dashed). There is no significant decrease in fractional count with increased $d_1$.}}
\label{fig:fracs}
\end{figure}


Fig.~\ref{fig:dist_asym_mr} replicates Fig.~\ref{fig:dist_asym_sm}, but with colours coded by the ratio of stellar masses of galaxies to those of their nearest neighbours. A Spearman rank correlation test between mass ratio ($\log(M_{*}/M_{\odot})_1-\log(M_{*}/M_{\odot})_2$ and asymmetry yields $\rho=-0.21, p=0.0029$, and the same test between mass ratio and distance gives $\rho=-0.16, p=0.026$. However, there is no correlation when the mass range is split into high and low mass samples ($\log(M_*/M_\odot)>10.0$, $\log(M_*/M_\odot)<9.0$). Of course, mass ratio and stellar mass are not independent. Indeed, a partial correlation test of mass ratio and asymmetry, taking stellar mass into account, yields $\rho=-0.016, p=0.78$, implying no independent correlation. Further, there is no restriction on $d_1$ when calculating the mass ratio, and galaxies with massive nearest neighbours at large $d_1$ are not likely to be perturbed.

Partial Pearson correlation tests for all galaxies in the sample of $d_{1}$, $\overline{v_{asym}}$ and stellar mass (Table~\ref{table:partial_corr}) show that $d_{1}$ and $\overline{v_{asym}}$ are independently correlated, as are $\overline{v_{asym}}$ and stellar mass. There is no significant independent relationship between $d_{1}$ and stellar mass. 
\begin{figure}
\centering
\includegraphics[width=9cm]{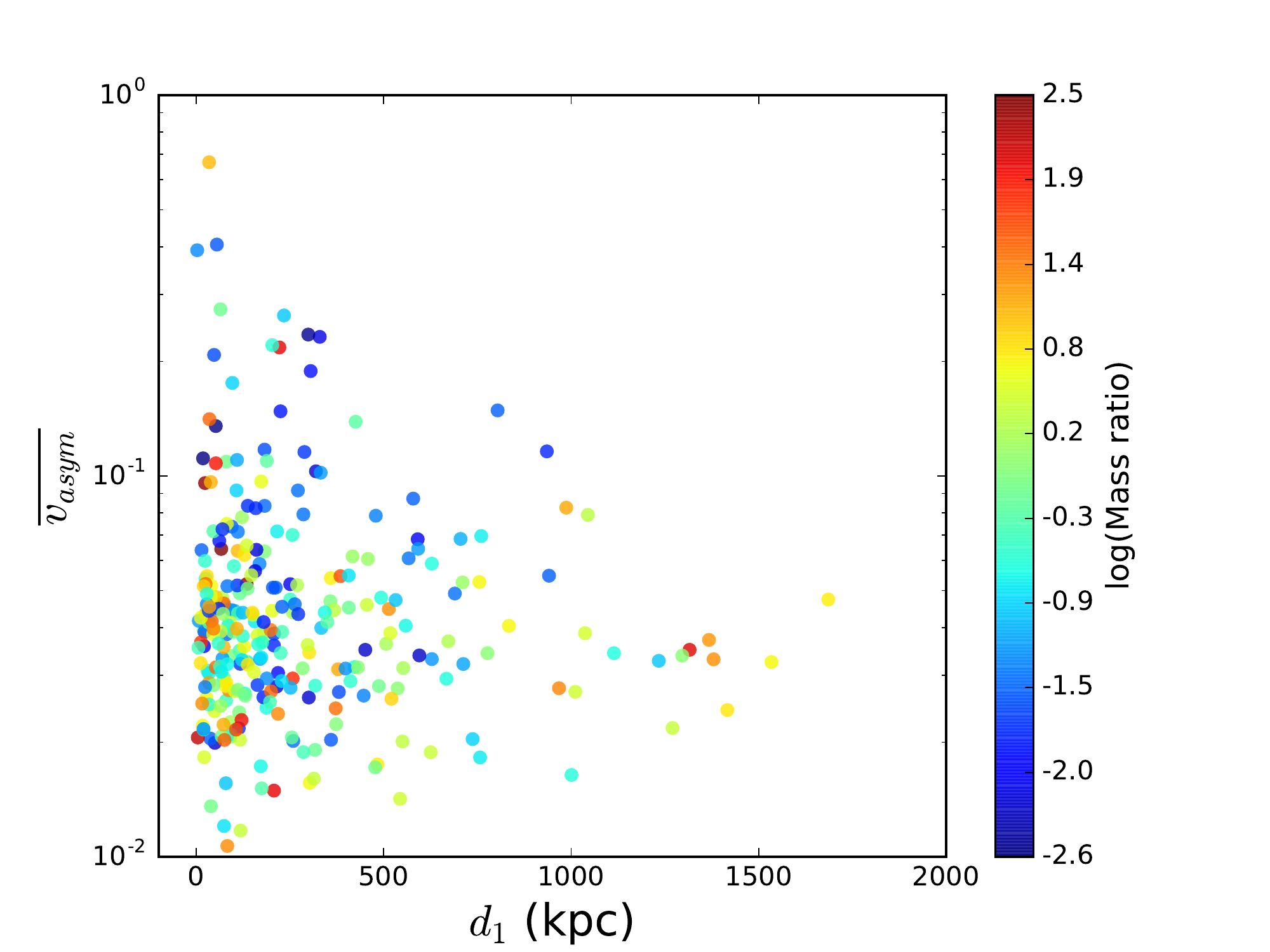}
\caption{$\overline{v_{asym}}$ against the comoving separation between galaxies and their (first) nearest neighbours, with points coloured by the ratio between the stellar masses of galaxies and their nearest neighbours. There is a relationship between mass ratio and $\overline{v_{asym}}$, but it appears to be dependent on the underlying relationship between mass ratio and stellar mass.}
\label{fig:dist_asym_mr}
\end{figure}

$d_1$ is, of course, only a proxy for interaction. In order to directly measure the gravitational influence exerted on galaxies by their nearest neighbour, Fig.~\ref{fig:dist_asym_force} shows $d_{1}$ against $\overline{v_{asym}}$ for galaxies in groups [identified using the GAMA Groups catalogue \citep{robotham2011galaxy}] with points coloured by the unitless tidal perturbation parameter from \citet{byrd1990tidal}:
\begin{equation}
P=\frac{M_{\mathrm{rat}}}{R^{3}}
\label{equation:p}
\end{equation}
where $M_{\mathrm{rat}}$ is the mass ratio described above and
\begin{equation}
R=\frac{d_{1}}{r_{90}}
\end{equation} 
where $r_{90}$ is the radius enclosing $90\%$ of the galaxy's light profile from the Sersic profile fits in the GAMA catalogue  \citep{kelvin2012galaxy}.
We only consider galaxies in groups (including pairs) because isolated galaxies, by definition, do not have nearby galaxies. The value for $P$ given here is the maximal value within a galaxy group, i.e. the value of the maximum tidal force exerted by a group member on the subject galaxy, as used in Schaefer et al. (submitted). There is no correlation between $P$ and $\overline{v_{asym}}$ for any mass or $d_1$ range within the sample. This holds true even when only considering galaxies with $P>0.1$, which is given by  \citet{byrd1990tidal} as the value above which tidal perturbation is likely to be effective. 




\begin{figure}
\centering
\includegraphics[width=9cm]{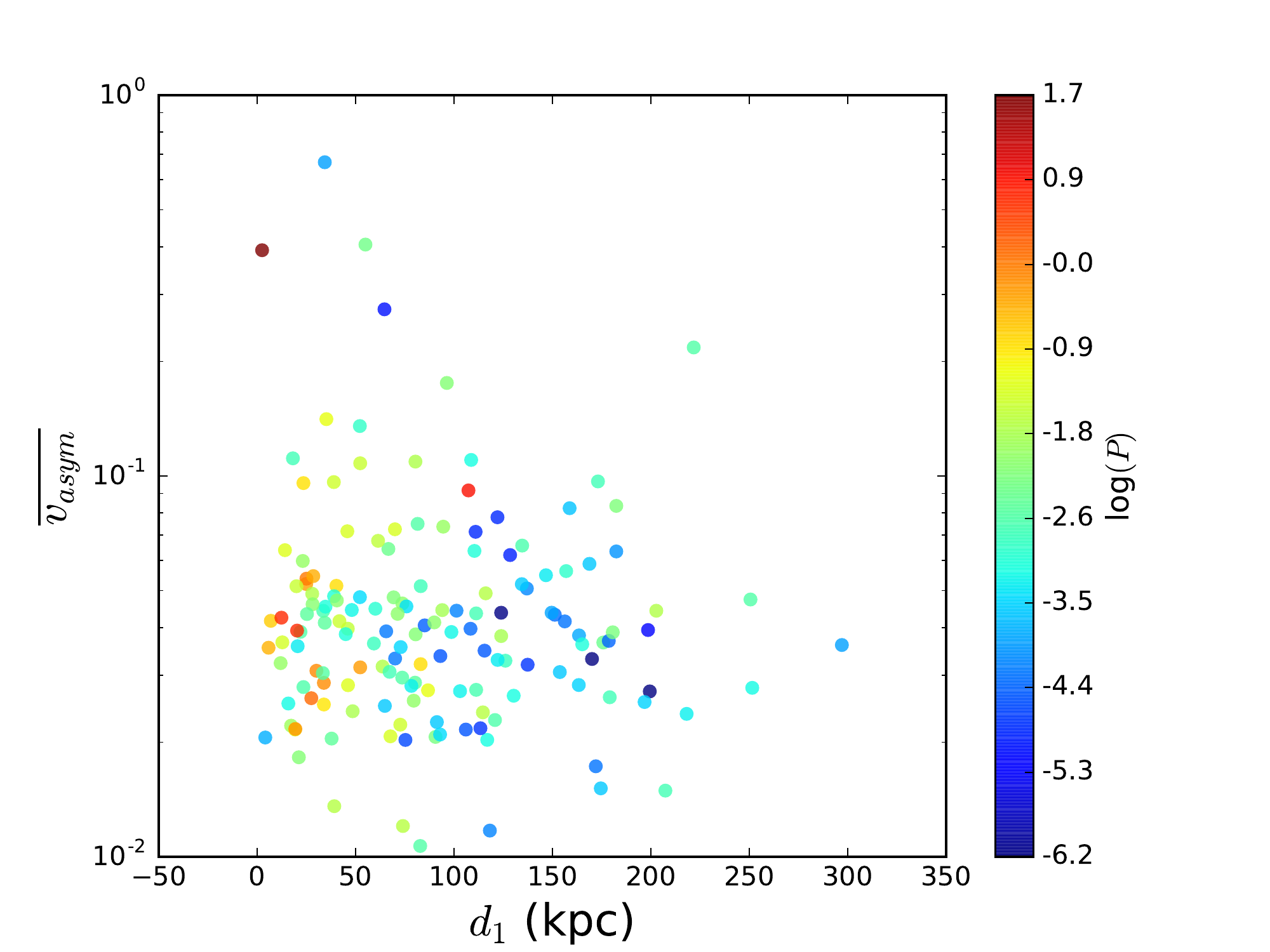}
\caption{$\overline{v_{asym}}$ against the comoving separation between galaxies and their (first) nearest neighbours, with points coloured by the tidal perturbation parameter $P$, as described in Equation~\ref{equation:p}. Of course, the strongest relationship is between $P$ (tidal force) and distance, and there is no correlation between $P$ and asymmetry.}
\label{fig:dist_asym_force}
\end{figure}

{In summary we find that, at all nearest neighbour separations, low mass galaxies have higher asymmetry than high mass galaxies, suggesting that processes other than interaction influence asymmetry for these objects. For high mass galaxies, asymmetry increases at low $d_1$, and there is an overall change in the distribution of asymmetries at low $d_1$, as seen in Fig.~\ref{fig:rms}. Low mass galaxies show only a weak change in the RMS of $\overline{v_{asym}}$ at different $d_1$, and there is no correlation between $d_1$ and $\overline{v_{asym}}$. This means that we cannot rule out a change in asymmetry in low mass galaxies at low $d_1$, but there is no strong signature of this effect.}


\section{Causes of kinematic asymmetry}
\label{sec:discussion}

The results in Section~\ref{sec:results} are indicative of plural causes of asymmetry in our sample. Further, there is evidence that there may be different primary causes of asymmetry for high and low mass galaxies. We here explore three potential causes of asymmetry, and how they affect galaxies with different stellar masses:
\begin{itemize}
\item interaction between galaxies, either through mergers or tidal disturbance
\item turbulence caused by star formation feedback and other processes linked to high gas fraction
\item asymmetric gas cloud distribution caused by low gas mass
\end{itemize}

\subsection{Interaction (pre- and post-merger)}
\label{sec:int}

Interaction is a known cause of kinematic asymmetry [e.g. \citet{krajnovic2006kinemetry,shapiro2008kinemetry}]. Whilst Fig.~\ref{fig:dist_asym_sm} shows no overall relationship between $d_1$ and $\overline{v_{asym}}$, Fig.~\ref{fig:dist_bins} shows that galaxies with $\log{M_*/M_\odot}>10.0$ show some evidence that asymmetry is correlated with distance to nearest neighbour. {The increased RMS at low $d_1$ across both high and low mass populations in Fig.~\ref{fig:rms} also indicates that interaction leads to high asymmetry values. That the dispersion increases at relatively large $d_1$ also suggests that it may not just be very local, galaxy-galaxy interactions that cause the environmental trends seen in the distribution of $\overline{v_{asym}}$.}


{In order to test whether asymmetry is caused by interaction, we use quantitative morphology measures Gini, $G$, and $M_{20}$, which can be combined to identify merging galaxies as in \citet{lotz2004new}}. The $G$, and $M_{20}$ coefficients are both measures of the distribution of light. $G$ is used to measure inequality in a distribution \citep{gini1912variabilita}, in this case in the spatial distribution of light \citet{abraham2003new}. High $G$ indicates high concentration of light, and $G$ is defined as \citep{glasser1962variance}:
\begin{equation}
G=\frac{1}{\overline{X}n(n-1)}\sum_{i}^{n}(2i-n-1)X_{i} ,
\end{equation}
where $\overline{X}$ is (in this case) the mean of the galaxy flux in $n$ total pixels, with $X_{i}$ being the flux in the $i$th pixel.

The $M_{20}$ coefficient is similar to $G$, in that it measures the concentration of light within a galaxy, but can be used to distinguish galaxies with different Hubble types \citep{lotz2004new}. The total second-order moment of galaxy flux, $M_{tot}$, is defined as the flux $f_{i}$ in each pixel, multiplied by the squared distance between pixel $i$ and the centre of the galaxy $(x_{c},y_{c})$, summed over all galaxy pixels:
\begin{equation}
M_{tot}=\sum_{i}^{n}M_{i}=\sum_{i}^{n}f_{i}[(x_{i}-x_{c})^{2}+(y_{i}-y_{c})^{2}] .
\end{equation}

$M_{20}$ is then defined as the normalised second-order moment of the brightest 20\% of the galaxy's flux. The galaxy pixels are rank-ordered by flux, and then $M_{i}$ (the second-order moment of light for each pixel $i$) is summed over the brightest pixels until the sum of the brightest pixels is equal to 20\% of the total flux:
\begin{equation}
M_{20}=log_{10}\left(\frac{\sum_{i}M_{i}}{M_{tot}}\right)
\end{equation}
for $\sum_{i}f_{i}<0.2f_{tot}$, where $f_{tot}$ is the total flux, $\sum_{i}^{n}f_{i}$. 

We use the $r$-band SDSS DR10 images \citep{ahn2014tenth} and follow \citet{lotz2004new} and \citet{conselice2008structures}. Fig.~\ref{fig:gm20} shows the $M_{20}$/Gini plane as in \citet{lotz2004new} for galaxies with $\log(M_*/M_\odot)>10.0$ and $<10.0$, respectively. The line delineating galaxies showing features associated with interaction suggested by \citet{lotz2004new} is shown in blue. Points are coloured by $\overline{v_{asym}}$. 

Whilst all galaxies above the line in both cases are asymmetric, in the high mass case there are no extremely asymmetric galaxies below the line. In the lower mass case, however, many of the galaxies with the highest value of $\overline{v_{asym}}$ lie below the line. Fig.~\ref{fig:gm20_diff} shows these differences in distribution more clearly. This may indicate that high mass galaxies with the highest asymmetry are interacting, whereas high asymmetry low mass galaxies may not necessarily be. Fig.~\ref{fig:23623} and Fig.~\ref{fig:618992} are examples of high mass galaxies above the line in the high mass panel in Fig.~\ref{fig:gm20}, both with highly disturbed velocity fields consistent with interaction. Fig.~\ref{fig:55160} is a low mass galaxy above the line in the right hand (low mass) panel of Fig.~\ref{fig:gm20}, and shows a kinematically decoupled core, signifying a potential past interaction. By contrast, Fig.~\ref{fig:560718} shows an asymmetric galaxy \emph{below} the line in the right hand panel of Fig.~\ref{fig:gm20}. We stress that the use of a cutoff mass of $\log(M_*/M_\odot)=10.0$ between the two planes in Fig.~\ref{fig:gm20} and Fig.~\ref{fig:gm20_diff} contributes to the observed `cleanness' of the difference between the distributions. Lower mass cuts reduce the distinction between the high and low mass samples, so the sharpness of the cutoff at $\log(M_*/M_\odot)=10.0$ should be interpreted with caution. 

\begin{figure*}
\centering
\includegraphics[width=8cm]{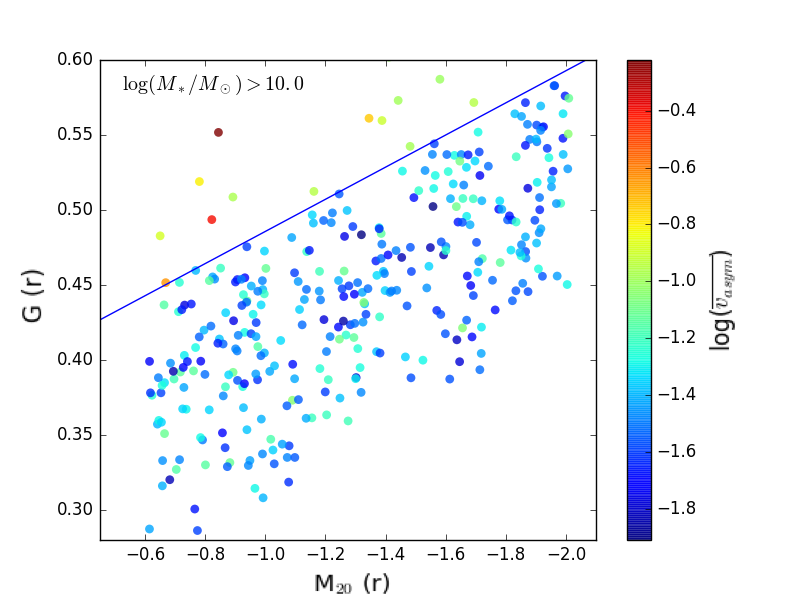} \includegraphics[width=8cm]{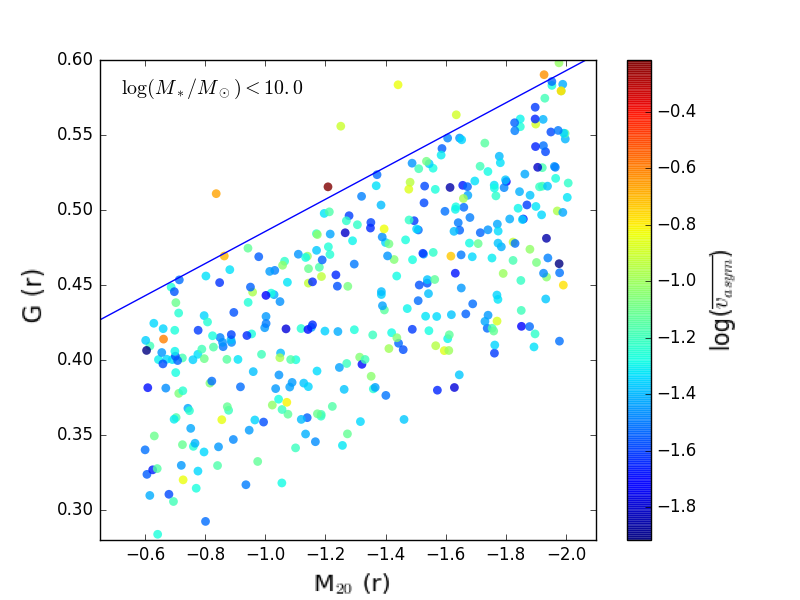}
\caption{The $M_{20}$/Gini plane from \citet{lotz2004new}, for galaxies with $\log(M_*/M_\odot)>10.0$ (left) and $<10.0$ (right), both with points coloured by $\overline{v_{asym}}$. The line delineating galaxies showing features associated with interaction from \citet{lotz2004new} is shown in blue. Whilst all galaxies above the line in both cases are asymmetric, in the high mass case there are no extremely asymmetric galaxies below the line. In the lower mass case, however, many of the galaxies with the highest value of $\overline{v_{asym}}$ lie below the line. This may indicate that high mass galaxies with the highest asymmetry are interacting, whereas high asymmetry low mass galaxies may not necessarily be.}
\label{fig:gm20}
\end{figure*}

\begin{figure*}
\centering
\includegraphics[width=8cm]{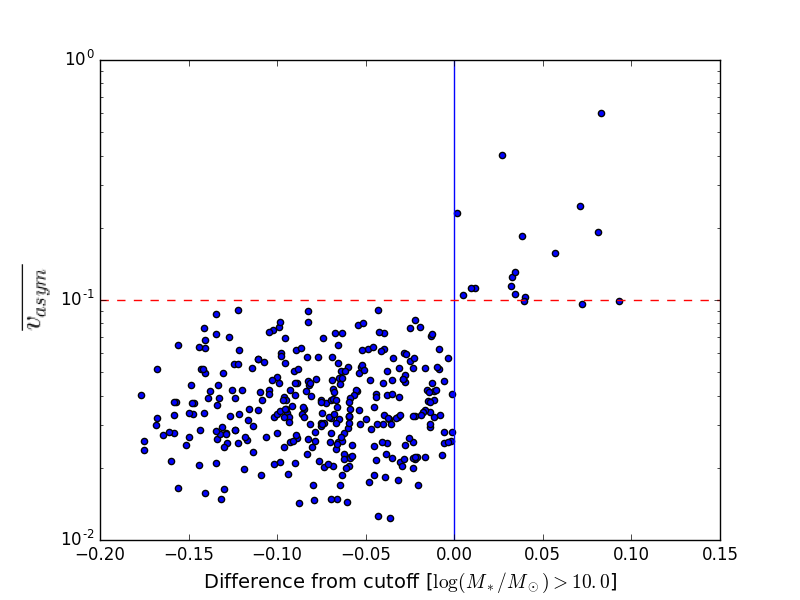} \includegraphics[width=8cm]{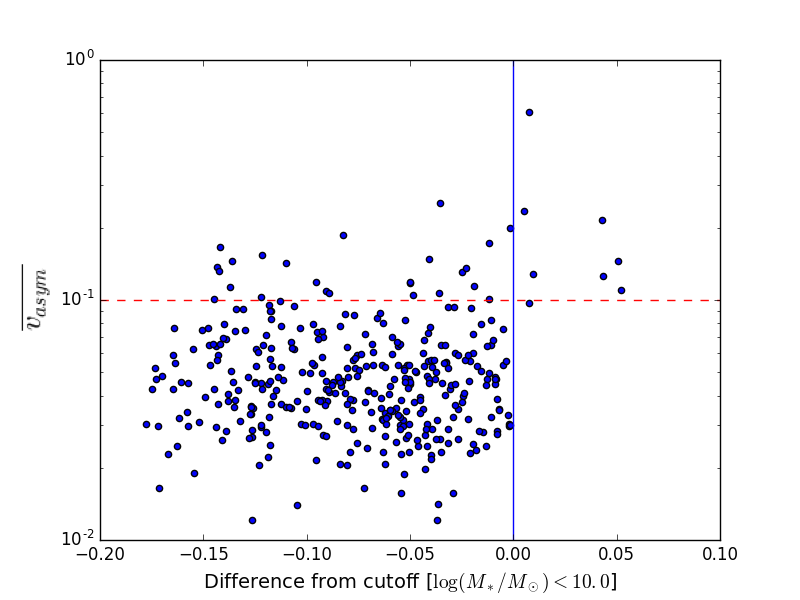}
\caption{Difference from the line on the $M_{20}$/Gini plane proposed by \citet{lotz2004new} (zero line shown in blue) against $\overline{v_{asym}}$, for galaxies with $\log(M_*/M_\odot)>10.0$ (left) and $<10.0$ (right). The red, dashed line shows $\overline{v_{asym}}=0.1$. In the high mass case, almost all galaxies with positive difference from the \citet{lotz2004new} line have $\overline{v_{asym}}>0.1$, and there are none of these very high asymmetry galaxies with negative residuals from the line. In the low mass case, whilst all galaxies with positive difference from the line have high asymmetry, a significant number of galaxies with $\overline{v_{asym}}>0.1$ show negative differences, i.e. they appear below the line in Fig.~\ref{fig:gm20}.}
\label{fig:gm20_diff}
\end{figure*}


\begin{figure}
\centering
\includegraphics[width=9cm]{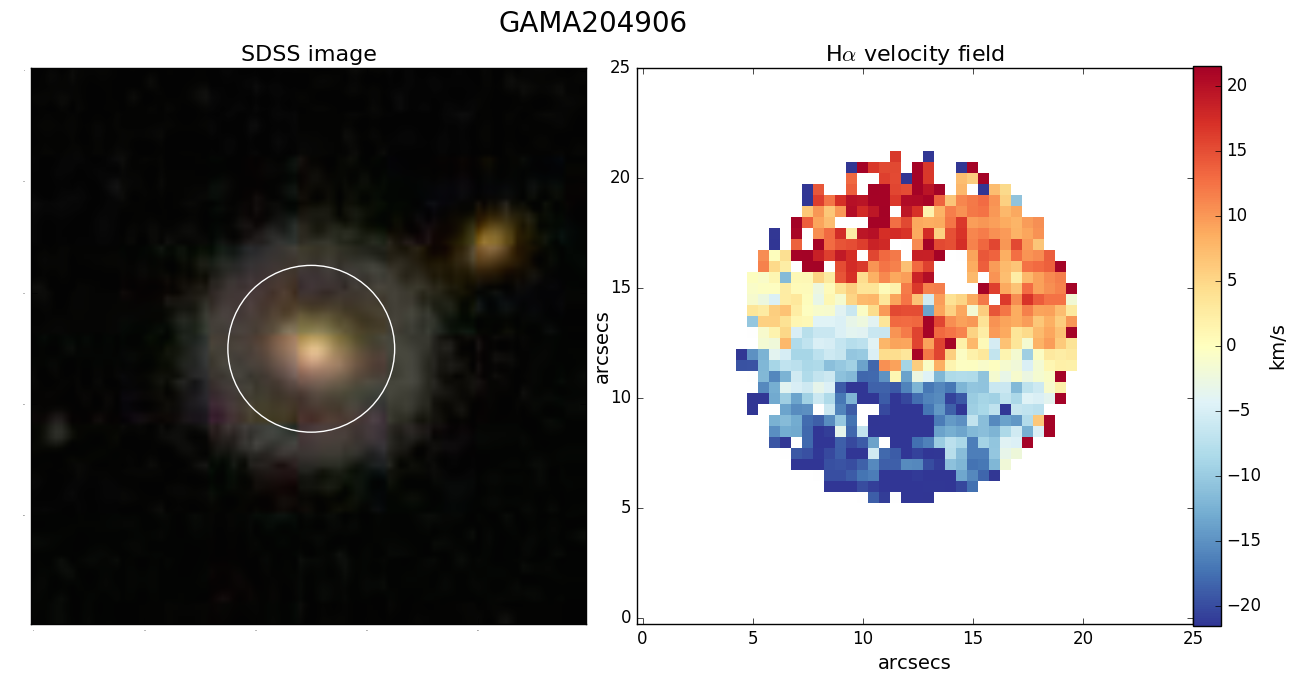}
\caption{The SDSS image (left) and H$\alpha$ velocity field (right) for GAMA204906, a galaxy with $\log(M_*/M_\odot)=10.4$ and $\overline{v_{asym}}=0.11$. It lies above the line from \citet{lotz2004new} in Fig.~\ref{fig:gm20}. The SAMI field of view is shown as a white circle.}
\label{fig:23623}
\end{figure}


\begin{figure}
\centering
\includegraphics[width=9cm]{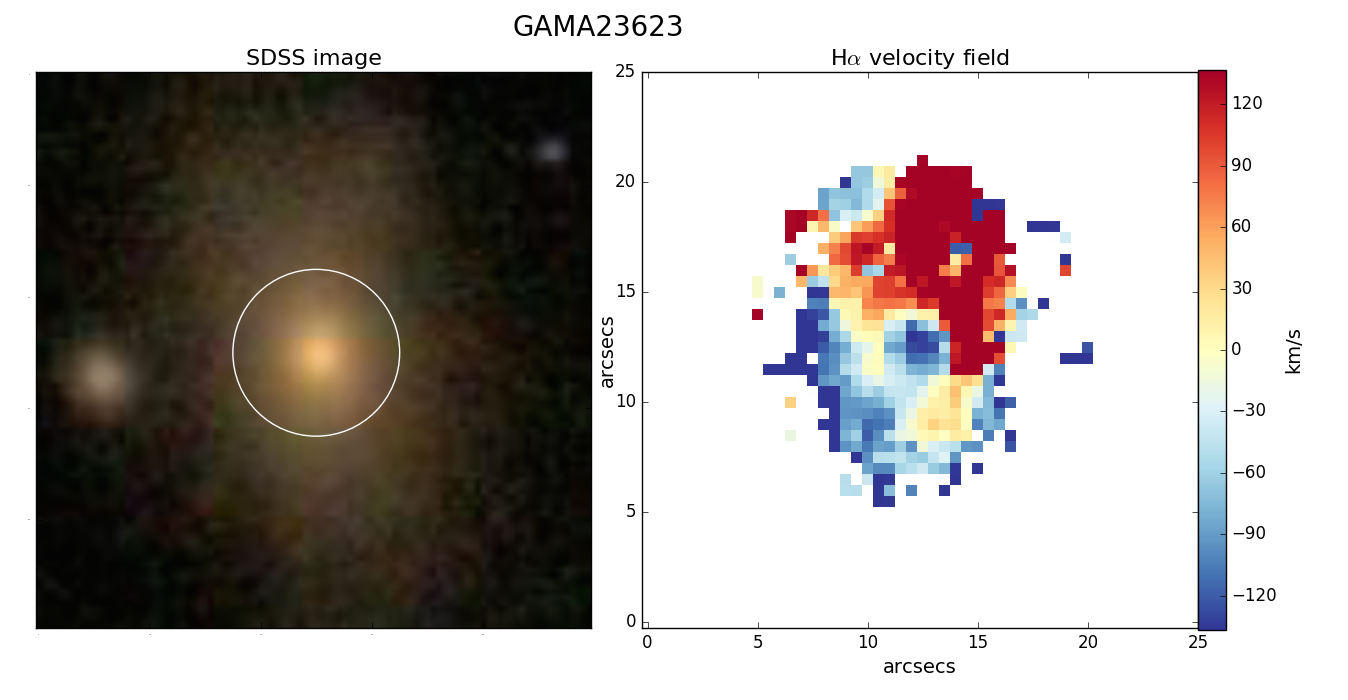}
\caption{GAMA23623, with $\log(M_*/M_\odot)=11.4$ and $\overline{v_{asym}}=0.19$. It lies above the line from \citet{lotz2004new} in Fig.~\ref{fig:gm20}.  The SAMI field of view is shown as a white circle.}
\label{fig:618992}
\end{figure}

\begin{figure}
\centering
\includegraphics[width=9cm]{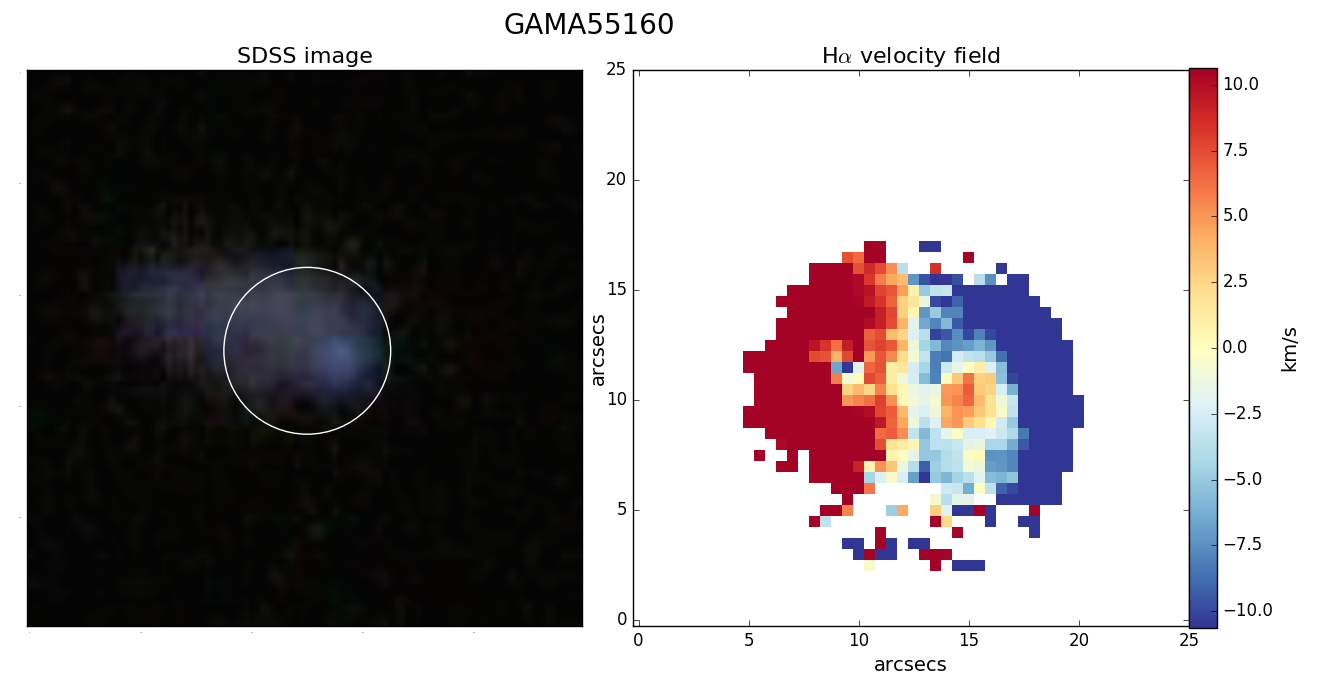}
\caption{GAMA55160, with $\log(M_*/M_\odot)=8.4$ and $\overline{v_{asym}}=0.14$. It lies above the line from \citet{lotz2004new} in Fig.~\ref{fig:gm20}. The kinematically decoupled core may be the result of an interaction.  The SAMI field of view is shown as a white circle}.
\label{fig:55160}
\end{figure}

\begin{figure}
\centering
\includegraphics[width=9cm]{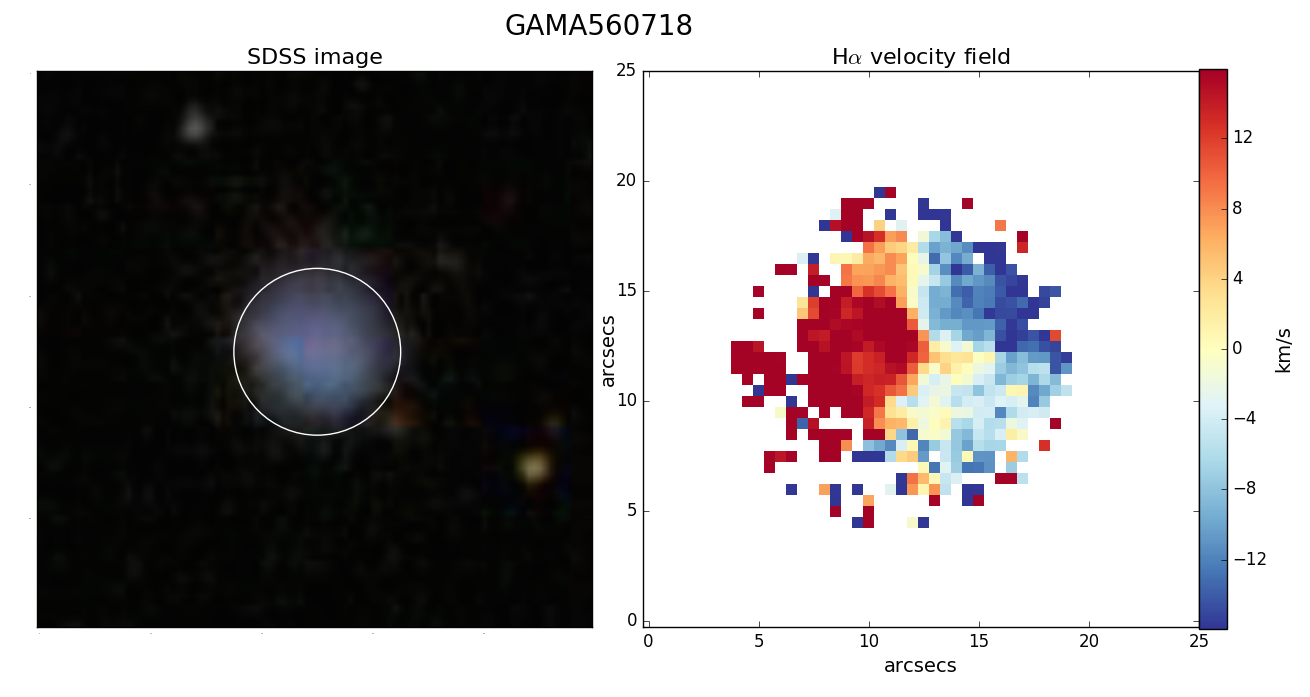}
\caption{GAMA560718, with $\log(M_*/M_\odot)=7.8$ and $\overline{v_{asym}}=0.10$. It lies below the line from \citet{lotz2004new} in Fig.~\ref{fig:gm20}, indicating that it may not show features of interaction.  The SAMI field of view is shown as a white circle.}
\label{fig:560718}
\end{figure}

The conclusion from the above and Fig.~\ref{fig:dist_bins} is that stellar mass and distance to nearest neighbour act separately on $\overline{v_{asym}}$, independently producing kinematic asymmetry in galaxies of different mass ranges. In the case of galaxies with $\log(M_*/M_\odot)>10.0$, interaction is a probable cause of asymmetry. However, whilst there will be some asymmetric, low mass galaxies that are either currently or have recently been interacting, this is not a sufficient explanation for the full range of asymmetry seen in the low mass sample.

It is also true that there is a large amount of scatter in all cases in Fig.~\ref{fig:dist_bins}. Whilst some galaxies in the $\log(M_*/M_\odot)$ can be definitively classified as interacting, and the observed correlation is statistically robust, it is also possible that other factors influence asymmetry in high mass galaxies, and that this contributes to the scatter in Fig.~\ref{fig:dist_bins}. 

In summary, all the statistical tests presented in this work suggest that there is a statistically significant correlation between $\overline{v_{asym}}$ and $d_1$, at least for galaxies with $\log(M_*/M_\odot)>10.0$. This suggests that galaxy-galaxy interactions are playing a role in driving the degree of asymmetry in at least a fraction of our sample. However, the large scatter present in Figs.~\ref{fig:dist_asym_sm}-\ref{fig:dist_asym_force} seems to imply that additional factors other than gravitational disturbances alone may also be responsible for the high value of $\overline{v_{asym}}$ observed in our sample. 




\subsection{Secular processes in low stellar mass galaxies}
\subsubsection{{Dispersion}}
\begin{center}
\begin{table}
  \begin{tabular}{ b{3cm}  l  b{1.5cm}  lb{1.5cm}  l }
    \hline
    x, y & $\rho$ & $p$ \\ \hline
    $d_{1},  \overline{v_{asym}}$ & -0.10 & 0.0046\\
    $d_{1},\log(M_{*}/M_\odot)$ & -0.067 & 0.060\\
    $\overline{v_{asym}}, \log(M_{*}/M_\odot) $ &  -0.18 & $3\times10^{-7}$\\
    \hline
  \end{tabular}
\caption{Partial Pearson correlation coefficients for stellar mass, $d_{1}$ and $\overline{v_{asym}}$.  $\overline{v_{asym}}$ and stellar mass remain correlated, ignoring $d_{1}$, and likewise for $\overline{v_{asym}}$ and $d_{1}$, ignoring stellar mass. There is no partial correlation between $d_{1}$ and stellar mass.} 
\label{table:partial_corr}
\end{table}
\end{center}

In our previous work \citep{bloom2016sami}, we postulated that the observed inverse correlation between stellar mass and kinematic asymmetry may be related to the higher gas fraction found in low stellar mass galaxies. Fig.~\ref{fig:gfrac} shows a positive correlation ($\rho=0.14, p=4\times10^{-4}$) between estimated gas fraction against $\overline{v_{asym}}$. As we do not have direct gas masses for our sample, {atomic} gas fractions are estimated from Equation 5 in \citet{cortese2011effect}:
\begin{equation}
\log\bigg(\frac{M(HI)}{M_{*}/M_{\odot}}\bigg)=-0.33(NUV-r)-0.40\log(\mu_{*})+3.37
\label{equation:gfrac}
\end{equation}
where $NUV-r$ is the $NUV-r$ magnitude and $\mu_{*}$ is the stellar mass surface density in $M_\odot \mathrm{kpc}^{-2}$. {We have checked these mass estimates against molecular masses calculated from the star formation rate from non-obscuration affected, sub-mm data as in \citet{federrath2017sami} for a sub-sample of our galaxies for which the requisite data were available. We note that we did not use the molecular masses for this work because Herschel fluxes were unavailable for the full mass range of our sample. Although there was scatter, the relationship between the mass estimates used here and those calculated from star formation rate was 1:1, implying that the systematic relationships found here would not change if we used the calculated masses.} As there is significant scatter in the relationship in Fig.~\ref{fig:gfrac}, we also show median asymmetries in bins of gas fraction. There is an upwards trend in the median asymmetry with gas fraction. The lowest gas fraction bin has a slightly higher asymmetry than the median trend, driven by the small population of galaxies with high $\overline{v_{asym}}$ and $\log(M_*/M_\odot)$ and low gas fraction.
\begin{figure}
\centering
\includegraphics[width=9cm]{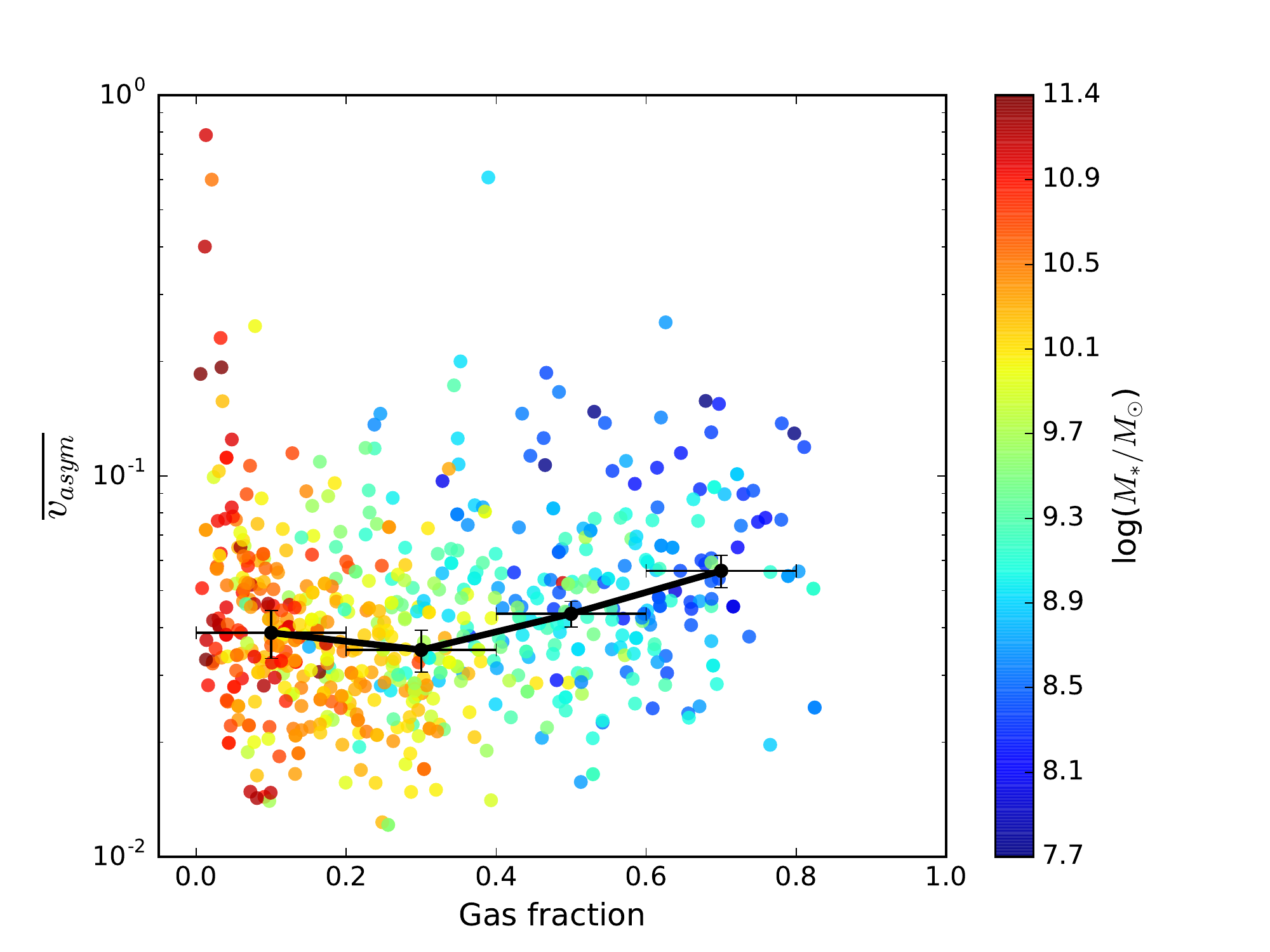}
\caption{Gas fraction against $\overline{v_{asym}}$ (on a log scale), showing a positive correlation. Points are coloured by $\log(M_*/M_\odot)$ This result supports the argument that asymmetry in our sample is linked to increased turbulence resulting from high gas fractions in low stellar mass galaxies. The expected relationship between gas fraction and $\log(M_*/M_\odot)$ is also clear, as determined by Equation~\ref{equation:gfrac}. Median asymmetries in bins of gas fraction are shown in black, with horizontal errors indicating bin width and vertical errors indicating the error on the median. There is an upwards trend in the median asymmetry with gas fraction. The first bin has increased $\overline{v_{asym}}$ caused by the interacting outlying galaxies with high $\overline{v_{asym}}$ and $\log(M_*/M_\odot)$ and low gas fraction.}
\label{fig:gfrac}
\end{figure}


\begin{figure}
\centering
\includegraphics[width=9cm]{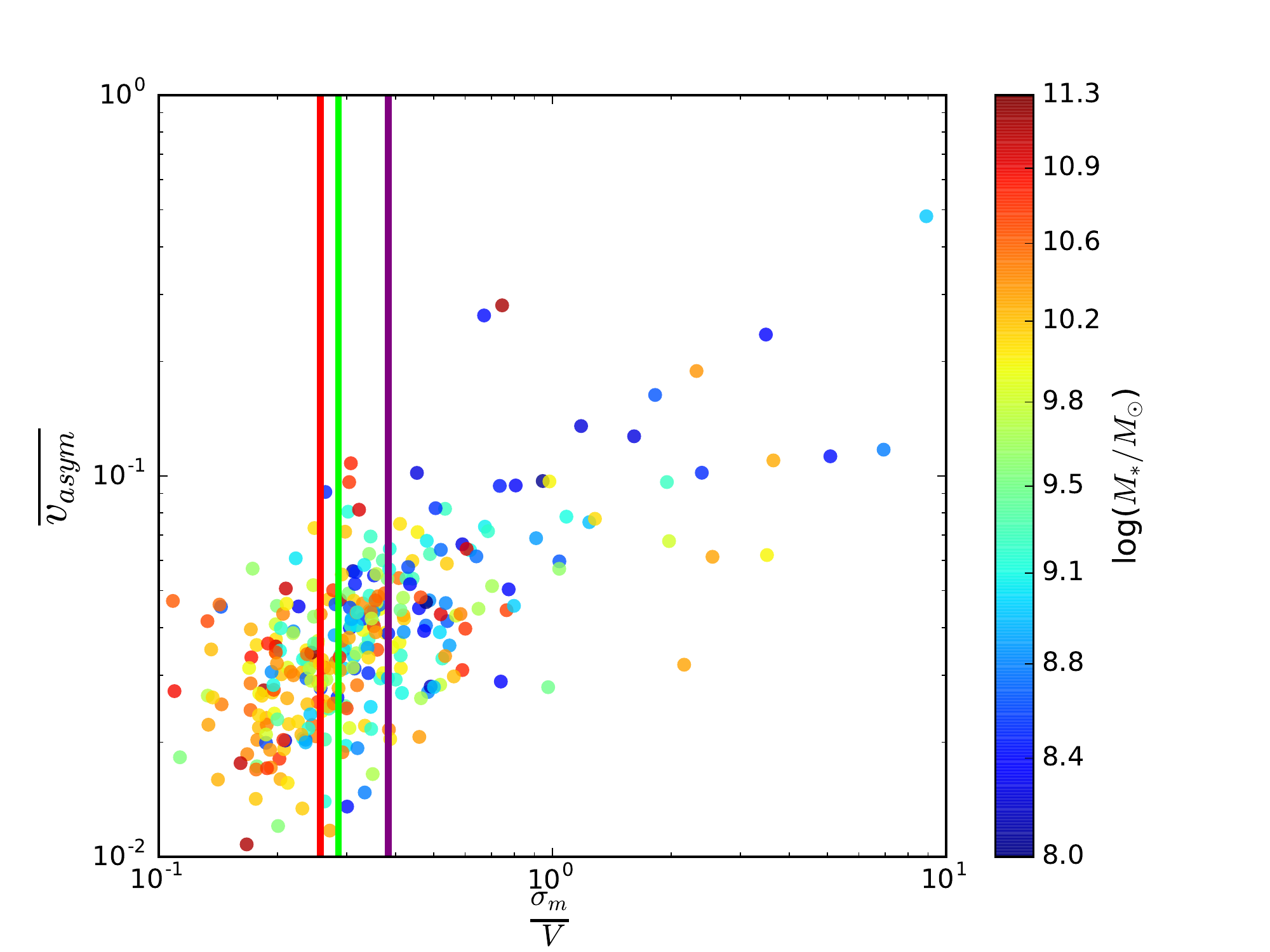}
\caption{$\frac{\sigma_{m}}{V}$ against $\overline{v_{asym}}$ (on a log scale), with points coloured by stellar mass. Median $\frac{\sigma_{m}}{V}$ for galaxies with $\log(M_{*}/M_\odot)<9, >9 $ and $>10.0$ are shown as purple, green and red, respectively. There is a significant offset in median $\frac{\sigma_{m}}{V}$ for galaxies with $\log(M_{*}/M_\odot)<9.0$. }
\label{fig:turbulence_1}
\end{figure}


The trend in Fig.~\ref{fig:gfrac} is in agreement with literature \citep{geha2006baryon,huang2012gas}. Work such as \citet{amorin2012complex} shows that complex H$\alpha$ kinematics in low mass galaxies are linked to the presence of multiple clumps of star formation. {We do, however, note that some systems, such as high-redshift sources, are asymmetric in flux but kinematically regular (e.g. \citet{wisnioski2011wigglez}).}

For star-forming galaxies, high gas fractions are linked to high  $\frac{\sigma_{m}}{V}$, where $\sigma_{m}$ is the mean gas velocity dispersion, and $V$ is the gas circular velocity \citep{green2013dynamo}. The $\sigma_{m}$ values we use are the mean dispersion from the SAMI velocity dispersion maps from {\small LZIFU}{\sc} (not flux-weighted). The $V$ values are taken from arctan fits to the radial $k_1$ profiles from the fitted kinemetry ellipses extracted as in Section~\ref{sec:kinemetry}. The value of the resulting fitted rotation curve is computed at $2.2r_e$ (where $r_e$ is the effective radius), as this is the point at which the curve is expected to {have become flat}. For a more thorough explanation of the fitting process, see \citet{bloom2017sami}.


Fig.~\ref{fig:turbulence_1} shows $\frac{\sigma_{m}}{V}$ against $\overline{v_{asym}}$ for all galaxies in the sample, with points coloured by stellar mass. Unsurprisingly, there is a strong relationship between $\frac{\sigma_{m}}{V}$ and $\overline{v_{asym}}$, as both measure {disturbances away from pure circular rotation}. A Spearman rank correlation test of $\frac{\sigma_{m}}{V}$ and  $\overline{v_{asym}}$ gives $\rho=0.56, p=6\times10^{-27}$.  Median $\frac{\sigma_{m}}{V}$ for galaxies with $\log(M_{*}/M_\odot)<9.0, >9.0 $ and $>10.0$ are shown as purple, green and red, respectively. Table~\ref{table:turb_meds} gives values for these medians. Low mass galaxies have significantly higher median $\frac{\sigma_{m}}{V}$ than higher mass galaxies. { This is to be expected given $V$ correlates with stellar mass.}

{In order to isolate the relationship between asymmetry and dispersion, without the dependence on stellar mass, we introduce the $T$ metric, which defines a measure of asymmetry as a fraction of dispersion}:
\begin{equation}
T=\overline{\left({\frac{k_3+k_5}{2\times k_0,\sigma}}\right)}
\label{equation:at_param}
\end{equation}
where $k_3$ and $k_5$, as in Section~\ref{sec:kinemetry} trace kinematic disturbance in the velocity field and $k_{0,\sigma}$ is, analogously to $k_1$, the radial values of zeroth order moment of the velocity dispersion field. This metric is an adaptation of the $\frac{\sigma_{m}}{V}$ parameter. We use $k_{0,\sigma}$ in this case, rather than $\sigma_{m}$, to increase ease of comparison to $\overline{v_{asym}}$.



Table~\ref{table:turb_meds} gives median values for $T$ for different mass ranges. Low stellar mass galaxies have significantly lower $\overline{T}$ i.e. they are more {dispersion-supported} for given values of $\overline{v_{asym}}$ than higher mass galaxies. There is also a positive Spearman rank correlation between $T$ and stellar mass, with $\rho=0.38, p=7\times{10^{-12}}$. Fig.~\ref{fig:ratmass} shows stellar mass against $T$, illustrating these relationships. Notably, the galaxies with the highest $T$ are at high stellar mass and, as discussed in Sec~\ref{sec:int}, high stellar mass galaxies with high $\overline{v_{asym}}$ show qualitative signs of interaction. This may point to $T$ as a means of distinguishing interacting galaxies from those with asymmetry caused by other processes. In order to confirm that the trend in Fig.~\ref{fig:ratmass} is not caused by the outliers with $T>0.2$, we show fit lines for the full sample and for the sample with those outliers excluded. Whilst the slope decreases slightly, the trend remains with the outliers excluded.

\begin{figure}
\centering
\includegraphics[width=9cm]{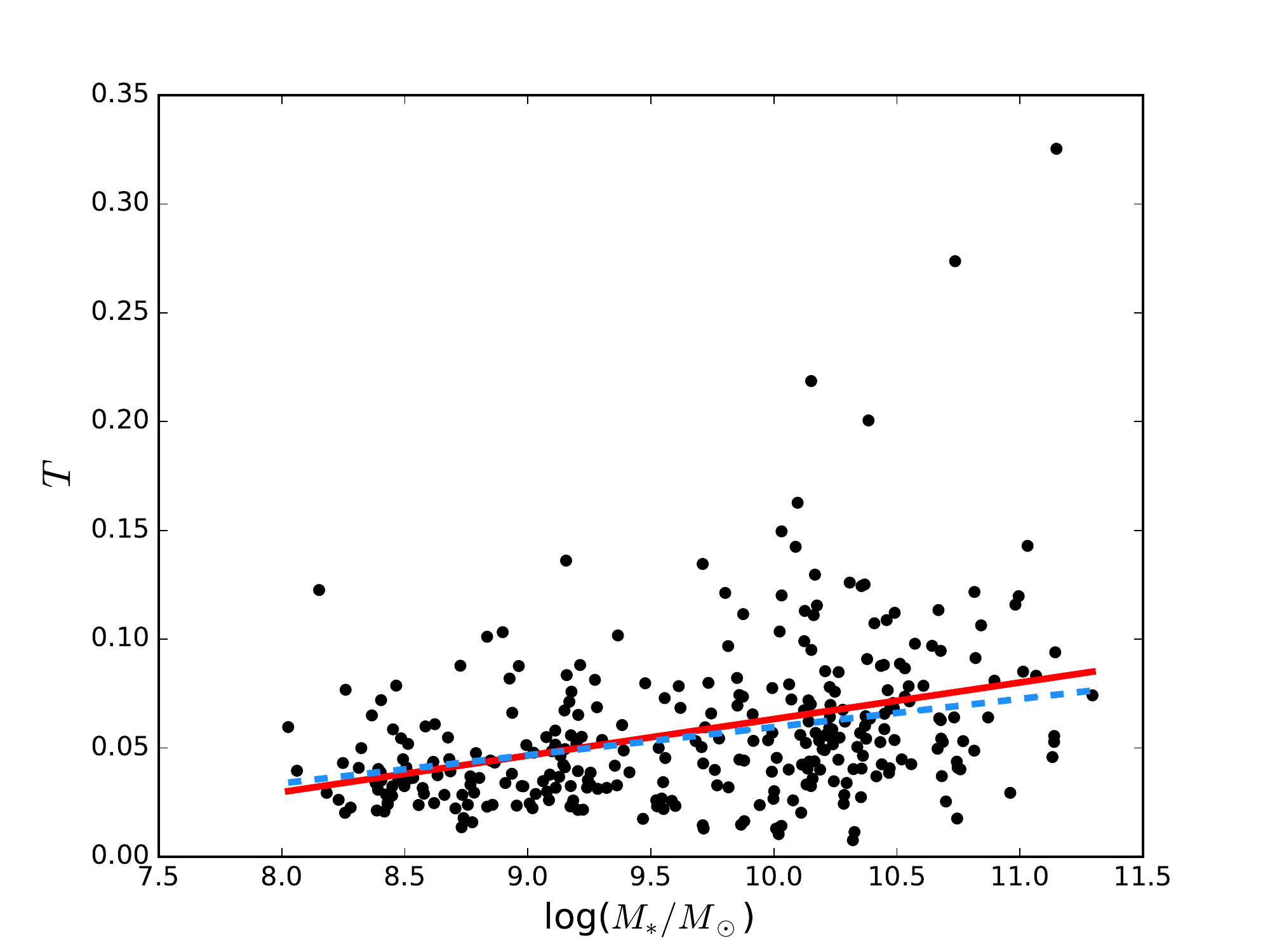}
\caption{Stellar mass against $T$ (defined in Equation~\ref{equation:at_param} as the ratio of the higher order moments of the velocity field to the velocity dispersion). There is a clear relationship, indicating that low stellar mass galaxies have lower velocity perturbation for given dispersions than high mass galaxies. For clarity, we also show the linear fit to the full sample (red) and the sample excluding the clear outliers with $T>0.2$ (blue, dashed). The trend remains with the outliers excluded.}
\label{fig:ratmass}
\end{figure}

The implication of these results is that low stellar mass galaxies have more small-scale than large-scale asymmetry than high stellar mass galaxies. Further, for given $\overline{v_{asym}}$, low stellar mass galaxies {have less fractional rotational support than their high mass counterparts}. This explains the results of Fig.~\ref{fig:dist_asym_mr} and Fig.~\ref{fig:dist_asym_force}, as neither mass ratio nor $P$ would be expected to correlate with asymmetry caused by {dispersion}.

\begin{center}
\begin{table}
  \begin{tabular}{ b{3cm}  l  b{2.0cm}  lb{1.0cm}   l }
    \hline
    Sample & $\overline{\frac{\sigma_{m}}{V}}$  & $\overline{T}$  \\ \hline
    $\log(M_{*}/M_\odot)<9.0$ & 0.38 $\pm$ 0.037 & 0.036 $\pm$ 0.0031\\
    $\log(M_{*}/M_\odot)>9.0$ & 0.28 $\pm 0.019$ & 0.054 $\pm$ 0.0033\\
    $\log(M_{*}/M_\odot)>10.0 $ &  0.25$\pm$ 0.024 & 0.062 $\pm$ 0.0048\\
    \hline
  \end{tabular}
\caption{Median $\frac{\sigma_{m}}{V}$ and median $T$ for galaxies with $\log(M_*/M_\odot)<9.0, >9.0 \ \& >10.0$, respectively. Low mass galaxies have both higher median $\frac{\sigma_{m}}{V}$ on the main sequence and lower median $T$, indicating that they are both more {dispersion-supported} overall and have higher {dispersion support} at a given $\overline{v_{asym}}$ than their higher mass counterparts.}
\label{table:turb_meds}
\end{table}
\end{center}

\subsubsection{Asymmetry of the gas cloud distribution}
\label{sec:joss}
Although low stellar mass galaxies have comparatively higher gas \emph{fractions}, they have lower overall gas \emph{masses}, which may affect the distribution of clouds of star-forming gas. Indeed, Fig.~\ref{fig:gfrac} shows that at high gas-fractions (i.e., $>0.4$) galaxies with lower stellar mass show larger asymmetry. This seems to suggest that - for dwarf galaxies - it is the gas mass (more than gas fraction relative to other baryonic components) that is linked to the degree of turbulence in the interstellar medium (ISM).   



Before discussing the physical implications of this result, it is important to remind the reader that what we plot are not measured cold gas fractions, but are empirical values. Taken at face-value, the $x$-axis in Fig.~\ref{fig:gfrac} should read as a combination of colour and stellar mass surface density, and the residual trends at high gas fractions as a likely secondary role played by either star formation rate and/or galaxy size. 

Since all galaxy properties are interconnected, disentangling between cause and effect does require direct cold gas measurements, which are unfortunately unavailable for our sample. However, it is natural to expect that the cold gas content is tightly connected to the degree of kinematic disturbance in the ISM. Thus, in the rest of the paper we assume that the Fig.~\ref{fig:gfrac} provides a fair representation of the real relation between gas content and $\overline{v_{asym}}$. It will be key for future works to further test this assumption. 

In order to explore the plausibility of gas mass as an influence on $\overline{v_{asym}}$, we create four simple models with different stellar masses ($\log(M_{*}/M_\odot)=8.0, 9.0, 10.0 $ and 11.0, respectively), evenly sampling the stellar mass range of our sample, to investigate the distribution of gas clouds in low and high mass galaxies and test this as an explanation for the observed asymmetry-stellar mass relationship. 


We divide the disc into 6 evenly spaced radial bins, after scaling all discs in terms of $r_e$ as in \citet{cecil2016sami}. The disc is further sub-divided into four azimuthal quadrants, to enable a calculation of spatial asymmetry. We then generate random populations of gas clouds according to several assumptions:
\begin{itemize}
\item the mass spectrum of clouds is the same across the disc, consistent with the IMF, as in \citet{bland2010long}. Given that the mass spectrum of star forming clouds is $10^3-10^7 M_\odot$, we take a median value $10^5 M_\odot$ weighted by IMF slope \citep{bland2010long}. There is some indication that cloud mass may scale with velocity dispersion and stellar mass surface density, as in \citet{krumholz2005general}, but this trend is likely to be comparatively weak.
\item that the HI gas fraction decreases monotonically with stellar mass. This assumption is broad, but serves to illustrate the general trend. We fit a line to the gas fractions estimated for our sample in Equation~\ref{equation:gfrac} and our stellar masses to determine the gas fraction for each of our model galaxies.
\item that the presence of gas clouds relates directly to star formation, one for one, through the Schmidt-Kennicutt law. In reality, there might be rich gaseous regions of the galaxy that have no star formation at a given phase in its evolution. The relation between star formation rate and gas surface density further depends on properties of the turbulence, in particular the Mach number \citep{federrath2013origin,salim2015universal}.
\item that all gas is in the form of clouds. This may not be a realistic assumption, but it is reasonable to assume that clouds will form at proportionately the same rate in high and low mass galaxies [supported by the cloud counts in \citet{bolatto2008resolved}], so changing the fraction of gas mass that is formed into clouds will not affect the relative results for galaxies of different masses.
\end{itemize}

We calculate the asymmetry in distribution of gas clouds by taking the maximal difference between counts of gas clouds in quadrants, normalised over the total number of gas clouds in the radial bin:
\begin{equation}
\mathrm{Gas\ cloud\ asymmetry=\frac{|N_{clouds,A}-N_{clouds,B}|}{N_{clouds,tot}}_{max}}
\end{equation}
where $\mathrm{N_{clouds,A}}$ and $\mathrm{N_{clouds,B}}$ are cloud counts in different quadrants and $\mathrm{N_{clouds,tot}}$ is the total cloud count in the radial bin.

Fig.~\ref{fig:joss} shows gas cloud asymmetry values for each of the six radial bins in each model galaxy, calculated for 1000 iterations of the random number generator. We observe a pronounced trend in which low mass galaxies have a more numerically asymmetric distribution of gas clouds, at all radii, than high mass galaxies.

\begin{figure}
\centering
\includegraphics[width=9cm]{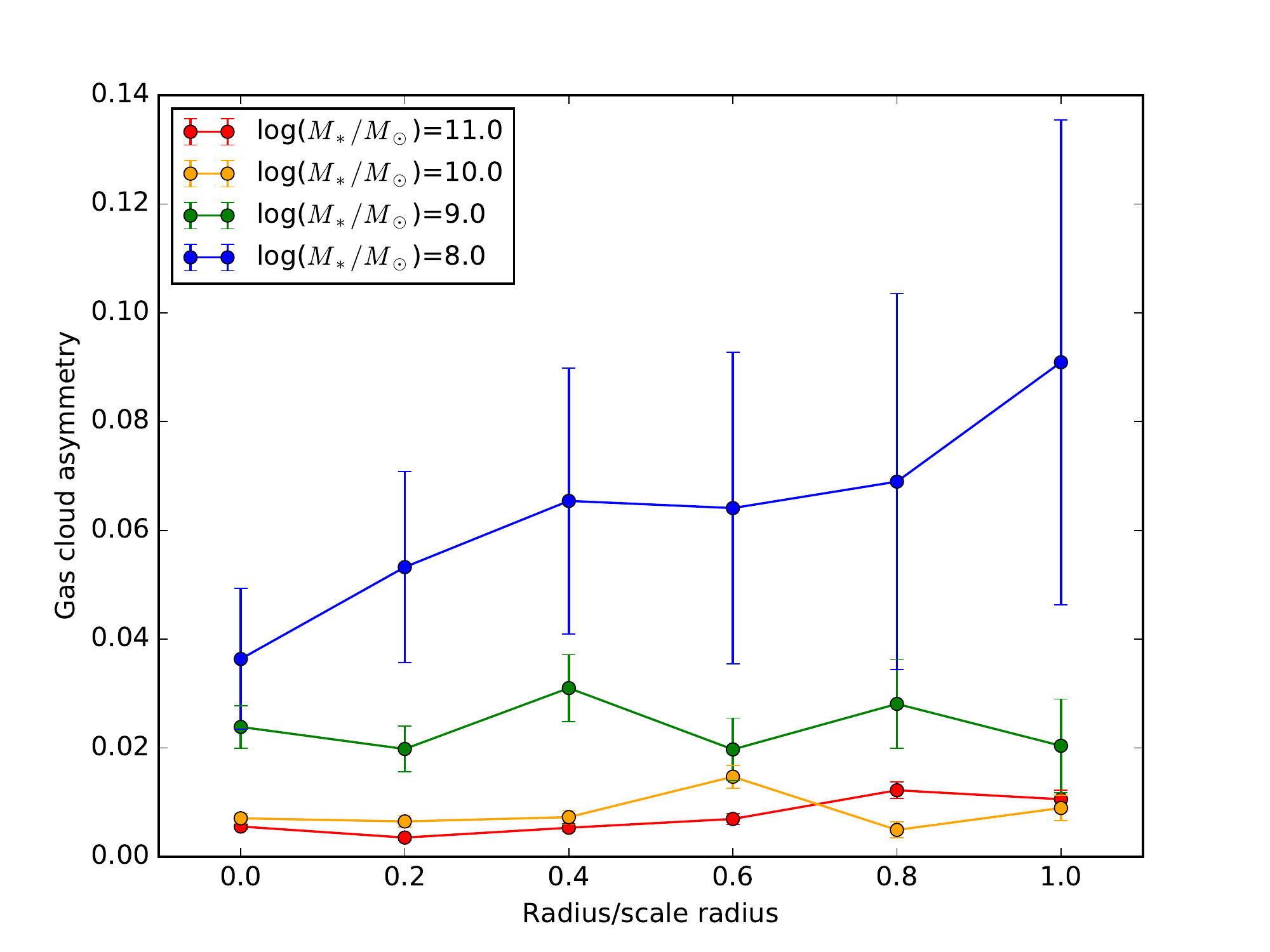}
\caption{Radius (normalised over scale radius) against gas cloud asymmetry for each of the four simple model galaxies. The gas cloud asymmetry is inversely proportional to stellar mass at all radii. Uncertainties are binomial 1$\sigma$  errors calculated following \citet{cameron2011estimation}.}
\label{fig:joss}
\end{figure}

Whilst this is a preliminary model, and does not incorporate many of the complexities which may affect gas cloud distribution, it illustrates the fundamental point that decreased gas mass leads to asymmetries in cloud distribution. This tendency towards `clumpy' star-forming regions may lead to the trends for low mass galaxies observed in the SAMI Galaxy Survey sample. \citet{ashley2013hi} show that morphological and kinematic asymmetry in low mass, low redshift galaxies are linked. They identify two galaxies from the Local  Irregulars  That  Trace  Luminosity  Extremes in The  HI Nearby  Galaxy  Survey (LITTLE THINGS) sample which have clumpy, asymmetrically distributed HI clouds and highly disturbed HI kinematics in corresponding regions. They proceed from the assumption that asymmetry is caused by interaction, and are thus surprised by the isolation of some disturbed, low mass galaxies in the LITTLE THINGS sample. Here, we speculate that the distribution of gas clouds in low gas mass galaxies provides a mechanism by which low mass galaxies may become perturbed in the absence of either ongoing or recent interactions. \citet{johnson2012stellar} also connect asymmetric clouds of HI gas with irregular HI rotation, and thence to disturbed H$\alpha$ morphology.

\citet{moiseev2012ionized} connect the magnitude of turbulent motion in dwarf galaxies to regions of high star formation, corresponding to gas clouds. They link local H$\alpha$ surface brightness to regions of high dispersion, such that maximum velocity dispersion is seen in areas of low H$\alpha$ surface brightness. That is, regions of high velocity dispersion are associated with areas surrounding regions of active star formation. \citet{moiseev2015controls} further argue that the velocity of turbulence in ionized gas is determined by star formation. Given that star formation is expected to be concentrated in giant molecular gas clouds \citep{blitz81giant}, it is plausible that the asymmetry of gas cloud distribution in Fig.~\ref{fig:joss} may lead to star formation in low mass galaxies being unevenly distributed over the disc. In low mass galaxies, the turbulence associated with star forming regions as in \citet{moiseev2012ionized} and \citet{moiseev2015controls} may be larger, relative to the underlying disc rotation, than in high mass galaxies.


The results and discussion presented here suggest several interesting directions for future work. Higher resolution kinematic maps, combined with imaging, of a statistical sample of local dwarf galaxies would enable the resolution of individual gas clouds and regions of star formation in both the kinematic and morphological regimes. Simulations with greater spatial resolution than SAMI may be able to provide the necessary data. We would then be able to trace the correspondence between asymmetric gas cloud distribution and kinematic disturbance more directly.  

Deep imaging would also provide a more reliable merger indicator than the  $M_{20}$/Gini plane. Whilst the  $M_{20}$/Gini plane may be used to identify a sample of galaxies which are most likely mergers, it is not clear that all mergers in the sample are identified. This would provide an important standard against which to calibrate $\overline{v_{asym}}$ values.

{Finally, it is possible that there is a relationship between star formation and $\overline{v_{asym}}$. \citet{bloom2016sami} found that there was no increase in global star formation for galaxies with $\overline{v_{asym}}>0.065$, but that there was a small increase in concentration of star formation. Further investigation of the links between concentration and the cloud geometry described here would be useful in determining whether star formation is causally linked to asymmetry.}

\section{Conclusions}
\label{sec:conclusion}
Whilst distance to first nearest neighbour, $d_{1}$, and thus interaction, plays some role in determining $\overline{v_{asym}}$, stellar mass is by far the strongest influence on kinematic asymmetry. Differences in relationships between $d_{1}$ and $\overline{v_{asym}}$, $\frac{\sigma_{m}}{V}$ and PA offset for high and low stellar mass galaxies point to qualitative differences between the samples.

We find no relationship between distance to fifth nearest neighbour and $\overline{v_{asym}}$.

For galaxies with $\log(M_{*}/M_\odot)>10.0$, there is an inverse relationship between $d_{1}$ and $\overline{v_{asym}}$, pointing to interactions as a source of kinematic asymmetry for high mass galaxies. Placement above the fiducial line on the $M_{20}$/Gini plane supports this.

However, for galaxies with $\log(M_*/M_\odot)<10.0$, a sample which includes many of the most asymmetric galaxies in our data set, there is no such relationship. We thus surmise that, for low stellar mass galaxies, there is a separate source of asymmetry beyond interaction.

Low stellar mass galaxies have higher {$\frac{\sigma_{m}}{V}$} than high mass galaxies. We also define a parameter $T$, representing asymmetry as a fraction of {dispersion}, and find that low stellar mass galaxies are more {dispersion-supported} for given values of asymmetry.

We propose two potential causes of increased turbulence in low stellar mass galaxies: increased gas fraction and decreased total gas mass. After preliminary investigation, gas mass appears to be the dominant influence, although further investigation of these effects will be the subject of future work. We suggest that higher resolution kinematic maps and deep imaging will be helpful in the future, in order to better trace the link between gas clouds and asymmetric gas kinematics.



\section*{Acknowledgements}

The SAMI Galaxy Survey is based on observations made at the Anglo-Australian Telescope. The Sydney-AAO Multi-object Integral field spectrograph (SAMI) was developed jointly by the University of Sydney and the Australian Astronomical Observatory. The SAMI input catalogue is based on data taken from the Sloan Digital Sky Survey, the GAMA Survey and the VST ATLAS Survey. The SAMI Galaxy Survey is funded by the Australian Research Council Centre of Excellence for All-sky Astrophysics (CAASTRO), through project number CE110001020, and other participating institutions. The SAMI Galaxy Survey website is http://sami-survey.org/.

GAMA is a joint European-Australasian project based around a spectroscopic campaign using the Anglo-Australian Telescope. The GAMA input catalogue is based on data taken from the Sloan Digital Sky Survey and the UKIRT Infrared Deep Sky Survey. Complementary imaging of the GAMA regions is being obtained by a number of independent survey programs including GALEX MIS, VST KiDS, VISTA VIKING, WISE, Herschel- ATLAS, GMRT    ASKAP providing UV to radio coverage. GAMA is funded by the STFC (UK), the ARC (Australia), the AAO, and the participating institutions. The GAMA website is http://www.gama-survey.org/.

Support for AMM is provided by NASA through Hubble Fellowship grant $\#$HST-HF2-51377 awarded by the Space Telescope Science Institute, which is operated by the Association of Universities for Research in Astronomy, Inc., for NASA, under contract NAS5-26555.

SMC acknowledges the support of an Australian Research Council Future Fellowship (FT100100457).

M.S.O. acknowledges the funding support from the Australian Research Council through a Future Fellowship (FT140100255).

We thank \'A.~R.~L\'opez-S\'anchez for assisting in the proofreading of the paper.

\bibliography{paper_II}
\bibliographystyle{mn2e}
\newpage
\appendix
\section{Correlation coefficients for variable mass ranges}
\label{sec:appendix}
{Fig:~\ref{fig:rp} shows Spearman rank correlation parameters $\rho$ and $p$ for the relationship between $\log(M_*/M_\odot)$ and $\overline{v_{asym}}$ for samples with variable ranges of stellar mass. The largest stellar mass range for which there is a (small) statistically significant Spearman rank correlation is $9.05<\log(M_*/M_\odot)<11.5$, with $\rho=-0.08, p=0.05$. The correlation increases in strength and decreases in $p$ with increasing minimum stellar mass. The smallest cutoff we consider is $10.5<\log(M_*/M_\odot)<11.5$ because there are not enough galaxies at high stellar mass for higher cutoffs.}

\begin{figure}
\centering
\includegraphics[width=8cm]{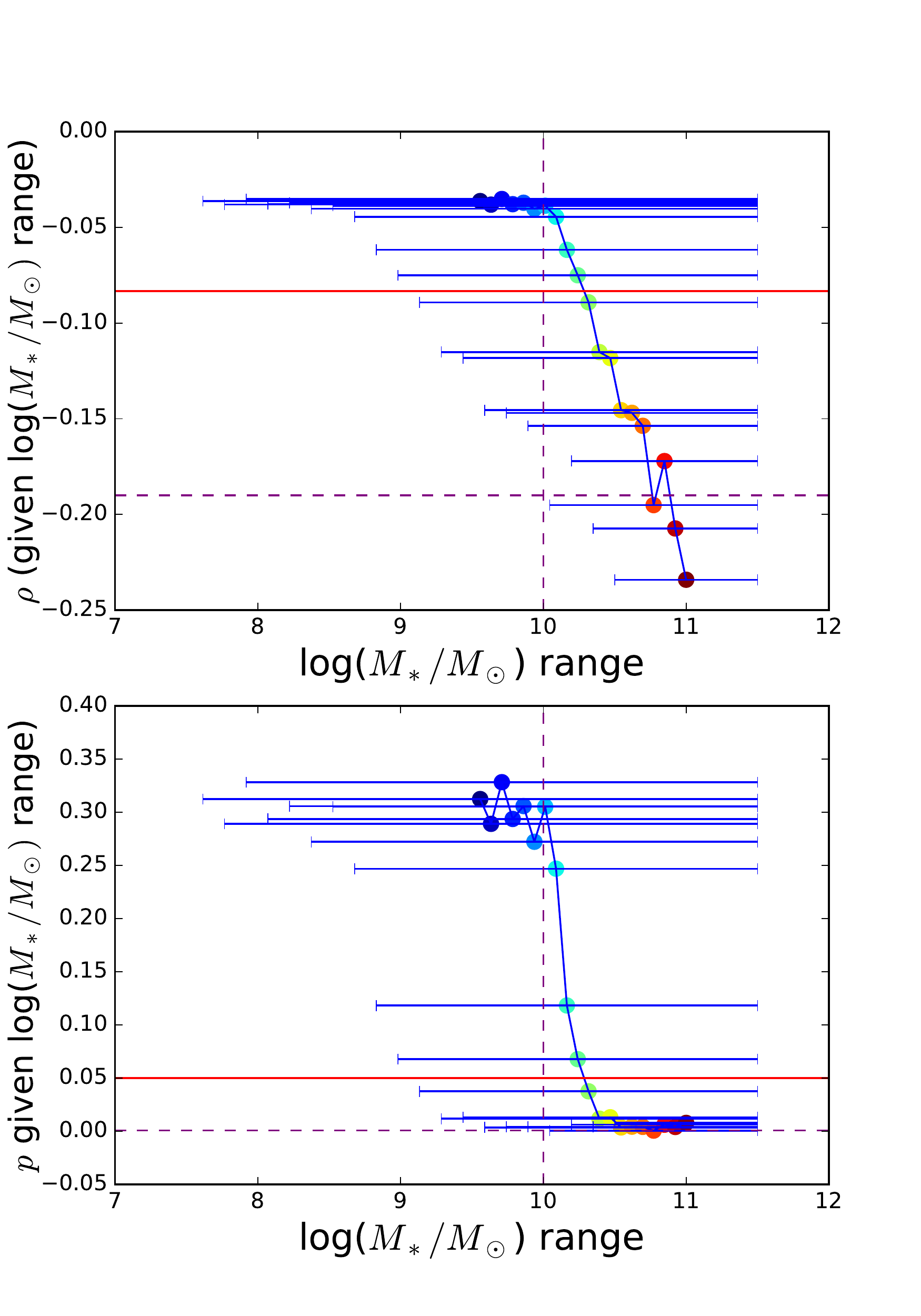}
\caption{Spearman rank correlation parameters $\rho$ and $p$ for the relationship between $\log(M_*/M_\odot)$ and $\overline{v_{asym}}$ for samples with variable stellar mass ranges, indicated by the error bars. For example, the purple dashed lines show the $\rho$ and $p$ for galaxies with $10<\log(M_*/M_\odot)<11.5$ ($\rho=-0.19, p=7\times10^{-4}$.). The red line indicates $p=0.05$ (and corresponding $\rho$ on the left panel). That is, there is a (small) statistically significant Spearman rank correlation for $9.05<\log(M_*/M_\odot)<11.5$, with $\rho=-0.08, p=0.05$. The correlation increases in strength and decreases in $p$ with increasing minimum stellar mass.}
\label{fig:rp}
\end{figure}

\end{document}